\documentclass[natbib]{emulateapj}       % for emulating 

\newcommand{\about}{\mbox{$\sim$}}               % ~ (about) 
\newcommand{\Av}{\mbox{$A_{V}$}}                  % Av
\newcommand{\frest}{\mbox{$f_{\rm rest}$}}      % f_rest
\newcommand{\Lbol}{\mbox{$L_{\rm bol}$}}                    % Lbol
\newcommand{\Lsol}{\mbox{$L_\odot$}}             % Lsol
\newcommand{\Lsun}{\mbox{$L_\odot$}}            % Lsun
\newcommand{\Mdyn}{\mbox{$M_{\rm dyn}$}}   % Mdyn
\newcommand{\Mmol}{\mbox{$M_{\rm mol}$}}    % Mmol
\newcommand{\Mbh}{\mbox{$M_{\rm bh}$}}        % Mbh
\newcommand{\Msol}{\mbox{$M_\odot$}}            % Msol
\newcommand{\Msun}{\mbox{$M_\odot$}}           % Msun
\newcommand{\NH}{\mbox{$N_{\rm H}$}}             % N_H
\newcommand{\ncrit}{\mbox{$n_{\rm crit}$}}         % ncrit
\newcommand{\nHH}{\mbox{$n_{\rm H_2}$}}       % n_H2
\newcommand{\signal}{\mbox{$\sigma$}}             % sigma (signal)
\newcommand{\Tb}{\mbox{$T_{\rm b}$}}               % Tb
\newcommand{\uv}{\mbox{$u$--$v$}}                     % u-v
\newcommand{\Vc}{\mbox{$V_{\rm c}$}}                % Vc
\newcommand{\Vsys}{\mbox{$V_{\rm sys}$}}        % Vsys
\newcommand{\asec}{\mbox{$''$}}                          % "
\newcommand{\hr}{\mbox{$^{\rm h}$}}                   % ^h
\newcommand{\mn}{\mbox{$^{\rm m}$}}                % ^m
\newcommand{\perMpc}{\mbox{Mpc$^{-1}$}}                              % Mpc^{-1}
\newcommand{\perbeam}{\mbox{beam$^{-1}$}}                         % beam^{-1}
\newcommand{\persquarecm}{\mbox{cm$^{-2}$}}                      % cm^{-2}
\newcommand{\percubiccm}{\mbox{cm$^{-3}$}}                         % cm^{-3}
\newcommand{\perHz}{\mbox{Hz$^{-1}$}}                                   % Hz^{-1}
\newcommand{\kms}{\mbox{km s$^{-1}$}}                                    % km/s
\newcommand{\persquarepc}{\mbox{pc$^{-2}$}}                         % pc^{-2}
\newcommand{\percubicpc}{\mbox{pc$^{-3}$}}                            % pc^{-3}
\newcommand{\peryr}{\mbox{yr$^{-1}$}}                                        % yr^{-1}
\newcommand{\peryear}{\mbox{yr$^{-1}$}}                                    % yr^{-1}
\newcommand{\unitofLM}{\mbox{$\Lsol \; \Msol^{-1}$}}                 % unit of L/M in solar
\newcommand{\unitofkappa}{\mbox{cm$^{2}$ g$^{-1}$}}         % unit of kappa
\newcommand{\unitofX}{\mbox{cm$^{-2}$ (K km s$^{-1}$)$^{-1}$}}   % unit of X
\newcommand{\plus}{\mbox{$+$}}    % +
\newcommand{\minus}{\mbox{$-$}}  % -
\newcommand{\plm}{\mbox{$\pm$}} % +- 
\newcommand{\twelveCO}{\mbox{$^{12}$CO}}                  % 12CO
\newcommand{\sixteenO}{\mbox{$^{16}$O}}                      % 16O
\newcommand{\eighteenO}{\mbox{$^{18}$O}}                    % 18O
\newcommand{\HH}{\mbox{H$_2$}}                                      % H2 (hydrogen molecule)
\newcommand{\water}{\mbox{H$_2$O}}                                      %
\newcommand{\HthirteenCN}{\mbox{H$^{13}$CN}}          % H13CN
\newcommand{\HCfifteenN}{\mbox{HC$^{15}$N}}             % HC15N
\newcommand{\HCthreeN}{\mbox{HC$_{3}$N}}                 % HC3N (cyanoacetylene)
\newcommand{\ammonia}{\mbox{NH$_3$}}                        % NH3
\newcommand{\HCOplus}{\mbox{HCO$^{+}$}}                    % HCO+(4-3)
\newcommand{\HCNv}{\mbox{HCN($v_2$=$1^{1f}$,J=4--3)}}      % HCN(v2=1^1f,J=4-3)
\newcommand{\NtwoHplus}{\mbox{N$_{2}$H$^{+}$}}       % N2H+
\newcommand{\nd}{\nodata}
\newcommand{\tnm}[1]{\tablenotemark{#1}}
\newcommand{\citest}[1]{\citeauthor*{#1}}
\newcommand{\citesp}[1]{(\citeauthor*{#1})}

\shorttitle{\apj\ in press}
\shortauthors{SAKAMOTO et al.}

\slugcomment{Accepted for publication in {\rm ApJ}}

\begin{document}
\title{Submillimeter Interferometry of the Luminous Infrared Galaxy NGC 4418: \\
A Hidden Hot Nucleus with an Inflow and an Outflow}

\author{Kazushi Sakamoto\altaffilmark{1}, 
Susanne Aalto\altaffilmark{2}, 
Francesco Costagliola\altaffilmark{2}, 
Sergio Mart\'{i}n\altaffilmark{3}, \\
Youichi Ohyama\altaffilmark{1},
Martina C. Wiedner\altaffilmark{4}, 
and
David J. Wilner\altaffilmark{5} \\
}
\altaffiltext{1}{Academia Sinica, Institute of Astronomy and Astrophysics, Taiwan} 
\altaffiltext{2}{Department of Earth and Space Sciences, Chalmers University of Technology, Onsala Space Observatory, Onsala, Sweden} 
\altaffiltext{3}{European Southern Observatory, Santiago, Chile}
\altaffiltext{4}{LERMA \& UMR 8112 du CNRS, Observatoire de Paris, 61 Av. de l'Observatoire,
75014, Paris, France}
\altaffiltext{5}{Harvard-Smithsonian Center for Astrophysics, Cambridge, MA, U.S.A.}

\begin{abstract}
We have observed the nucleus of the nearby luminous infrared galaxy NGC 4418
with subarcsec resolution at 860 and 450 \micron\ for the first time
to characterize its hidden power source.
A \about20 pc (0\farcs1) hot dusty core was found inside a 100 pc scale concentration of molecular gas 
at the galactic center.
The 860 \micron\ continuum  core has a deconvolved (peak) brightness temperature of 120--210 K.
The CO(3--2) peak brightness temperature there is as high as 90 K at 50 pc resolution.
The core has a bolometric luminosity of about $10^{11}$ \Lsol, which accounts for most of the galaxy luminosity.
It is  Compton thick ($\NH \gtrsim 10^{25} \; \persquarecm$)
and has a high luminosity-to-mass ratio \about500 \unitofLM\ 
as well as a high luminosity surface density $10^{8.5\pm 0.5} \; \Lsol\ \persquarepc$. 
These parameters are consistent with an AGN to be the main luminosity source (with an Eddington ratio about 0.3)
while they can be also due to a young starburst near its maximum $L/M$.
We also found an optical color (reddening)  feature that we attribute to an outflow cone emanating from the nucleus.
The hidden hot nucleus thus shows evidence of both an inflow, previously seen with absorption lines, 
and the new outflow reported here in a different direction.
The nucleus must be rapidly evolving with these gas flows.
\end{abstract}


\section{Introduction}  
\label{s.introduction}
There are galaxies hiding a compact luminous source in their nuclei
and \object{NGC 4418} is their local prototype. 
The extremely extinguished power source in this early-type disk galaxy was first recognized by \citet{Roche86}
who singled out the galaxy for its infrared brightness.
They found very deep silicate absorption at 9.7 \micron\ toward the infrared-bright nucleus
and deduced that the source of the infrared luminosity (\about$10^{11}$\Lsol)
is completely obscured by a dust shroud of \Av\ $\gg$ 50 mag. 
They further inferred from the weakness of line features in the infrared and optical spectra that 
the luminosity source is an accreting massive black hole (i.e., active galactic nucleus or AGN), 
although they also mentioned a nascent starburst embedded within a dusty cocoon as a possible alternative.

Subsequent observations to be reviewed in \S\ref{s.n4418review} have supported 
the presence of a deeply buried compact luminosity source, although its nature is still debated.
NGC 4418 is the most absorption-dominated galaxy in the mid-infrared
diagnostic diagram of \citet{Spoon07}. 
Its location on the diagonal branch in the diagram suggests that 
the luminosity source is almost completely covered by opaque dust. 
Usual spectroscopic signs of AGN can be completely extinguished by this heavy absorption.
Hence, the main piece of evidence for a hidden AGN, among other indirect ones, has been 
the compactness of the source inferred from the absorption depth. 
It has been argued 
that dust is unlikely to cover an extended starburst so completely as to
cause the deep absorption features \citep{Roche86,Dudley97,Spoon01}.
On the other hand, the arguments for very young starburst are based on the low radio to far infrared ratio 
and the detection of a warm molecular gas absorber toward the nucleus.
The former ratio is lower than in radio-quiet AGNs and is consistent with a starburst in the pre-supernova age 
($<$ 5 Myr) \citep{Roussel03}. 
The absorbing molecules appear to be in the region where X-rays would
have destroyed them if the heating is due to an AGN \citep{Lahuis07}.

The true nature of the luminous nucleus is important 
because this type of heavily extinguished nuclei 
could double the number of local AGN \citep{Maiolino03} 
or 
it could represent a young phase of unusually intense and compact starburst ($\ll$ 100 pc).
Luminous hidden nuclei of this kind are also important because
they may be in a short transition phase to unobscured AGNs, i.e.,
in the process of removing the obscuring gas \citep[e.g., through an AGN-driven outflow,][]{Sanders88,Fischer10}.
Such feedback is regarded as a key process in galaxy evolution and hence requires detailed study 
\citep{Hopkins06,Weiner09}.

A new approach  to characterize and diagnose the compact, luminous, and extinguished 
nuclei of this kind is  through submillimeter-wave observations at high spatial resolution. 
Sub-mm observations directly probe the thermal dust emission that 
constitutes most of the bolometric luminosity of such a nucleus.
Unlike optical and infrared observations, one can look deep into the obscuring material 
since dust opacity decreases at longer wavelengths.
Radio interferometry is available in the sub-mm to achieve sub-arcsec resolution that is currently 
unavailable in far-IR where the radiation of such a nucleus peaks.
It is possible to estimate the bolometric luminosity and luminosity density of the
nucleus using the Stefan-Boltzmann law by measuring the source size and brightness temperature 
at  a frequency where the nucleus has a photosphere.
One can also probe the nucleus by measuring its mass from rotation of circumnuclear molecular gas, 
which can be observed using sub-mm molecular lines.  
The luminosity-to-mass ratio, \Lbol/\Mdyn, so obtained will further constrain the nature
of the nucleus since an AGN can have a higher $L/M$ than stars.
Molecular lines can provide additional information on temperature, density,
chemical composition of the interstellar medium (ISM), and on the
radial motion of the gas affecting the evolution of the dusty shroud.
We had applied this approach to the ultraluminous infrared galaxy \object{Arp 220}, 
which shares some of the above mentioned characteristics of NGC 4418 including the deep silicate absorption.
We found that the brighter one of the twin nuclei of Arp 220 has \Lbol $\geq 2\times 10^{11}$ \Lsol\ and
\Lbol/\Mdyn $\gtrsim 400$ \Lsol/\Msol\ in the central 80 pc \citep[hereafter S08]{Sakamoto08}
and that each nucleus has a molecular outflow causing molecular P-Cygni profiles \citep{Sakamoto09}.

In this paper, we apply the same approach to NGC 4418 to quantitatively constrain the nature of the
deeply extinguished nucleus through sub-mm high-resolution observations.
We made the first sub-arcsecond resolution imaging of the galaxy in the submillimeter band,
observing 860 \micron\ continuum and molecular lines including CO(3--2), HCN(4--3), and \HCOplus(4--3)
and also imaging the nucleus with 450 \micron\ continuum.
We also analyzed optical multi-color images to probe circumnuclear dust distribution for signs of disturbance
and feedback.
The obtained parameters of the nucleus  are similar to the ones in Arp 220.
They are consistent with any of the following being energetically dominant:
an AGN, 
a young and compact starburst, 
and an AGN-compact starburst combination.
Our results support the previous studies using an independent method and
indicate a possible path to break the degeneracy about the luminosity source  
with future observations of higher quality.
We also found a sign of outflow from the nucleus in the optical data.
In the following, we briefly review properties of NGC 4418 and previous studies on the hidden nucleus 
in \S \ref{s.n4418review}.
We describe our SMA observations in \S \ref{s.obs} and present the results in \S \ref{s.result},
where we also report our examination of optical images of the galaxy.
The results are discussed in  \S \ref{s.discussion} and the conclusions are summarized in \S \ref{s.summary}.
Our companion paper reports high-resolution 1 mm and cm-wave observations of the same galaxy
\citep[hereafter C12]{Costagliola12}.

%%%%%%%%%%%%%%%%%%%%%%%%%%%%%%
\subsection{NGC 4418} 
\label{s.n4418review}

\begin{figure}[t]
\epsscale{0.45}
\plotone{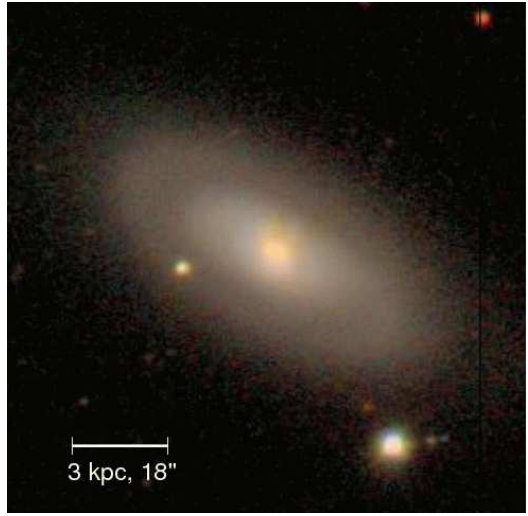}  % compressed for arXive   %{f01.rgb.eps}  
\caption{ \label{f.sdss}
NGC 4418 optical image from the Sloan Digital Sky Survey (0.3--1 \micron\ composite).
The image size is 100\arcsec\ (16.5 kpc) on a side.
North is up and east is to the left.
}
\end{figure}

Studies on NGC 4418 and its nucleus since \citet{Roche86} are summarized below.
(To guide readers, an optical image and key parameters of the galaxy are 
in Fig. \ref{f.sdss} and Table \ref{t.4418param}, respectively.
We adopt the galaxy distance of 34 Mpc at which 1\arcsec\ is 165 pc.)
(1) A high infrared-to-optical luminosity ratio $L_{\rm IR}/L_{\rm B}\approx 20$ \citep{Armus87,Young89}
and the shape of the spectral energy distribution (SED) suggest that most of the luminosity of the galaxy
is absorbed and reradiated by dust as thermal emission \citep{Roche93}. 
(2) NGC 4418 has warm mid-IR and far-IR colors
$S(25\micron)/S(60\micron)=0.220$ and $S(60\micron)/S(100\micron)=1.37$ \citep{Sanders03}. 
The former puts the galaxy in the `warm' population that often host active nuclei \citep{deGrijp85,Sanders88}.
The latter also suggests warm dust (color temperature of \about50 K) indicative of reprocessed dust emission. 
(3) The deep absorption has been confirmed by further spectroscopy and modeling; 
the estimated opacity is $\tau_{9.7 \; \mu m} \approx 7$ or \Av\ \about\ 100 mag 
\citep{Roche86,Dudley97,Spoon01,Siebenmorgen08}.
A radial gradient of temperature in the nucleus may also contribute to the deep 10 \micron\ depression 
in the same way \citet{Kwan76} modeled for protostars \citep{Dudley97}.
It is also suggested from a shallow SED slope in the FIR  
that the dust has a high enough column density to be optically thick 
even at or beyond 100 \micron\ \citep{Roche93,Lisenfeld00}.
Likely related to this is an unusually low ratio of the [\ion{C}{2}] 158 \micron\ line to far-IR
luminosity in NGC 4418 \citep{Malhotra97,Gracia-Caprio11}.
(4) The compactness of the infrared nucleus has been directly confirmed with mid-IR imaging
at subarcsecond resolutions \citep{Wynn-Williams93,Evans03,Siebenmorgen08}.
According to the Keck imaging by \citet{Evans03}, the 10 \micron\ size of the nucleus is 0\farcs3 
and the 25 \micron\ size is less than 0\farcs6, which is the diffraction limit of the 10 m telescope.
(5) The nucleus is also compact (\about0\farcs1--0\farcs4) and has a high brightness temperature
($\gtrsim 10^{5}$ K) at \about2 GHz 
(\citealt{Condon90,Kewley00,Baan06}). 
Such a radio nucleus is either an active nucleus or a compact starburst or their hybrid \citep{Smith98}.
(6) Chandra X-ray observations ``may imply the presence of a Compton-thick AGN'' although the identification is 
``somewhat tentative'' because of the limited photon statistics \citep{Maiolino03}.
(7) NGC 4418 has far-IR excess compared to its radio strength, 
attributed either to a dust-enshrouded AGN or to a young, compact starburst \citep{Kawara90,Yun01,Roussel03}.
(8) Molecular gas concentration to the central $\leq$ 5\arcsec\ has been observed with interferometers 
(\citealt{Imanishi04,Dale05,Sakamoto10}; \citest{Costagliola12}).
The mass of the molecular gas will be about $1  \times 10^{9}$ \Msol\  
if the average conversion factor in our Galaxy \citep{Hunter97} 
applies to the observed CO(1--0) flux of $100\pm50$ Jy \kms\ \citep{Kawara90,Sanders91,Young95,Dale05,Albrecht07}. 
NGC 4418 has a high infrared luminosity-to-molecular gas mass ratio of
$L_{\rm IR}/\Mmol \sim 100\; \Lsol/\Msol$ for the above \Mmol\ \citep{Kawara90,Sanders91}.
High-excitation molecular lines with excitation temperatures above 100 K have been
detected toward the nucleus 
(\citealt{Aalto07b,Lahuis07,Costagliola10,Sakamoto10,Gonzalez-Alfonso12}; \citest{Costagliola12}). 
The last four publications report vibrationally excited molecules and 
the last three report lines involving energy levels as high as \about1000 K.
(9) Redshifted absorption lines have been found toward the nucleus
by \citet{Gonzalez-Alfonso12} in the mid-infrared \ion{O}{1} and OH lines
and by \citet{Costagliola12} in the \ion{H}{1} 21 cm line. 
These are attributed to a gas inflow.

%%%%%%%%%%%%%%%%%%%%%%%%%%%%%%%%%%%%%%%%%%%%%%%%%%%%%%%%%%%%
\section{Submillimeter Array Observations} 
\label{s.obs}
\subsection{350 GHz (860 \micron) Observations}
We observed NGC 4418 in 2009, 2010, and 2012 at around 350 GHz (860 \micron) using 
the Submillimeter Array (SMA)\footnote{
The Submillimeter Array is a joint
project between the Smithsonian Astrophysical Observatory and the
Academia Sinica Institute of Astronomy and Astrophysics, and is
funded by the Smithsonian Institution and the Academia Sinica.
}.  We used the subcompact (SC), extended (EXT), and very extended (VEX) array configurations.
Observational parameters are in Table \ref{t.obslog}.
The data from the SC configuration were already reported in \citet[hereafter S10]{Sakamoto10}
and are combined here with the new data.
The overall coverage of baseline lengths is 6--509 m or 7--590 k$\lambda$ at 350 GHz for the galaxy.
We simultaneously observed CO(3--2) and \HCOplus(4--3) lines
in 2009 and CO(3--2) and HCN(4--3) in 2010 using 2 GHz of bandwidth. 
We used in the EXT and SC configurations
4 GHz bandwidth to simultaneously observe CO(3--2), \HCOplus(4--3), and HCN(4--3).
The SMA primary beam covers about 34\arcsec\ (5.6 kpc) in full width at half maximum (FWHM) 
at our observing frequencies.
The SMA spectrocorrelator was configured to the resolution of 3.25 MHz (about 3 \kms) and recorded
data from both upper and lower sidebands (USB and LSB).
We used 3C273 for our gain, passband, and pointing calibrations.
We also observed the quasar J1222+042 and Titan in our 2010 VEX observations for 
further calibration and its verification. 
Flux calibration was made using Titan and Ganymede in 2009, and Titan in 2010 and 2012.
The angular distances from our gain calibrator  3C273 to NGC 4418, J1222+042, 
and Titan in our 2010 observations, were only 3.0, 2.7, and 4.0 degrees, respectively. 
This mitigated calibration errors due to direction-dependent
phase de-correlation, atmospheric transmission, and pointing offsets.

We reduced our data using MIR for calibration \citep{Scoville93}, 
MIRIAD for imaging \citep{Sault95}, and AIPS \citep{Bridle94} and stand-alone
programs for further data analyses.
The data were first imaged at 30 \kms\ resolution for the entire bandwidths to identify lines.
Then the `line-free' channels were identified and integrated to create continuum data.
That continuum is then subtracted from the original to make continuum-subtracted visibilities.
The spectra and the identified lines are presented in \S \ref{s.result.spectra}.
The line data were further analyzed after binning to 10--30 \kms\ resolutions.
We did self-calibration of phase using the continuum data when making
images from the VEX data, although we did not do that when measuring the source position or
fitting the visibilities to determine the size and flux density of the continuum source. 
Table \ref{t.data} summarizes the properties of the images and spectral data cubes used in this paper.

%%%%%%
\subsection{660 GHz (450 \micron) Observations}
\label{s.660obs}
We also observed the galaxy at 660 GHz (450 \micron) during our VEX 350 GHz observations in 2009 and 2010
by allocating 2 GHz of correlator bandwidth to our high frequency receivers.
Only the 2010 run resulted in useable data from five antennas. 
Atmospheric absorption was significant at our observing frequency
even though the weather condition was excellent for Mauna Kea.
The mean zenith opacity was 0.038 at 225 GHz and 0.89 at 350 \micron\ during our observations
according to the monitoring at the adjacent Caltech Submillimeter Observatory. 
The opacity should be about the same at 450 and 350 \micron.
The median double sideband (DSB) system temperature was about 1200 K toward NGC 4418.
We used 3C273 for both passband and gain calibration.   
Both 350 GHz and 660 GHz data were used for the gain calibration.
We first applied the 350 GHz gain solution with its phase multiplied 
by the ratio of the observing frequencies, 1.9, and then derived another 
gain solution using the 660 GHz data and applied it to improve the calibration. 
No strong line is in our passband and the data are treated as continuum.
Since Titan is too heavily resolved, flux calibration at 660 GHz is 
based on the 350 GHz flux density of 3C273 scaled to 660 GHz using
the spectral index of $\alpha=-1.0$ \citesp{Sakamoto10}.
We estimate our flux measurements at 660 GHz to be accurate only to 36\%   
(for $\sigma(\alpha)$ = 0.3 and 30\% error for other calibrations).

%%%%%%%%%%%%%%%%%%%%%%%%%%%%%%%%%%%%%%%%%%%%%%%%%%%%%%%%%%%
\section{Observational Results and Analysis}
\label{s.result}

%%%%%%%%%%%%%%%%%%%%%%%%%
\subsection{Spectra and Line-Continuum Separation}
\label{s.result.spectra}

\begin{figure}[t]
\epsscale{0.75}
\plottwo{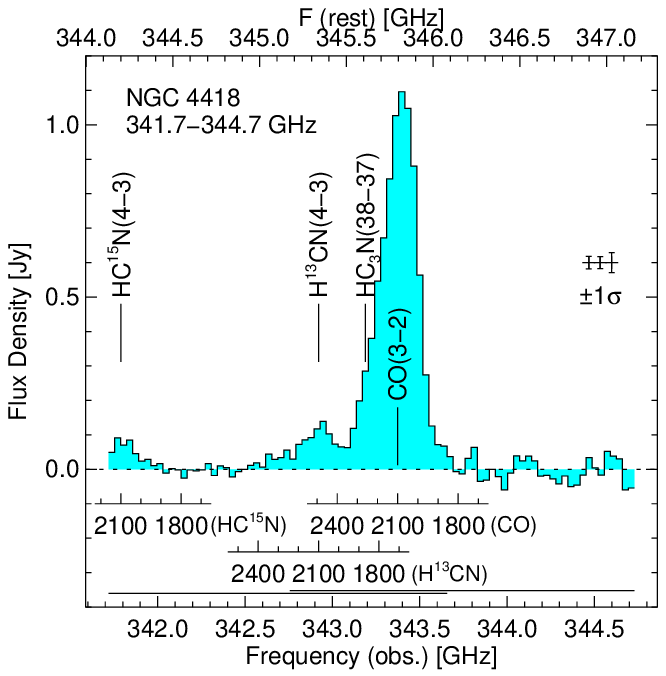}{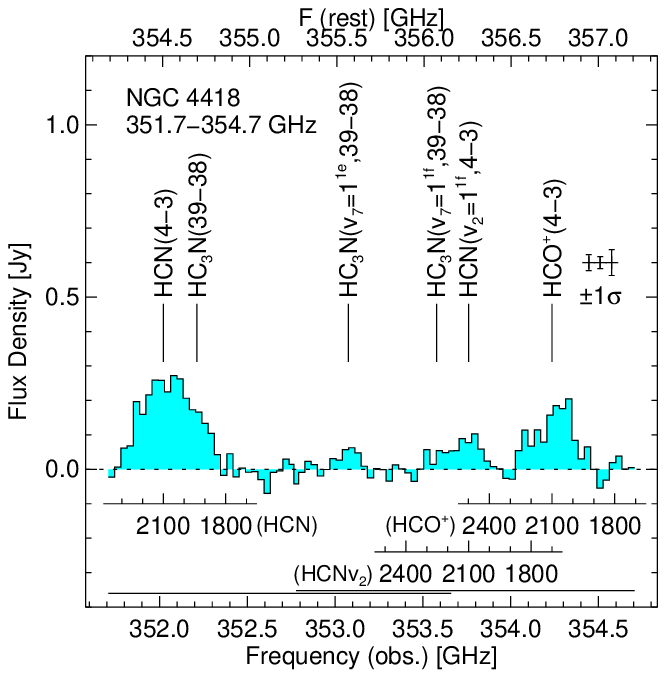}
\epsscale{1.0}
\caption{ \label{f.345VEXspec}
NGC 4418 spectra in the central 0\farcs5 (80 pc; FWHM). 
Data from the VEX array configuration are used and continuum has been subtracted.
Emission lines are marked at the redshift of 2100 \kms\ and six of them have velocity axes.
The data are from two observations with offset frequency coverages indicated with 
the two horizontal lines at the bottom. 
The data at the overlapped frequencies are averaged.
The three error bars in each panel are for the three spectral sections (e.g., the middle one for the 
averaged section).
}
\end{figure}

\begin{figure}[h]
\plottwo{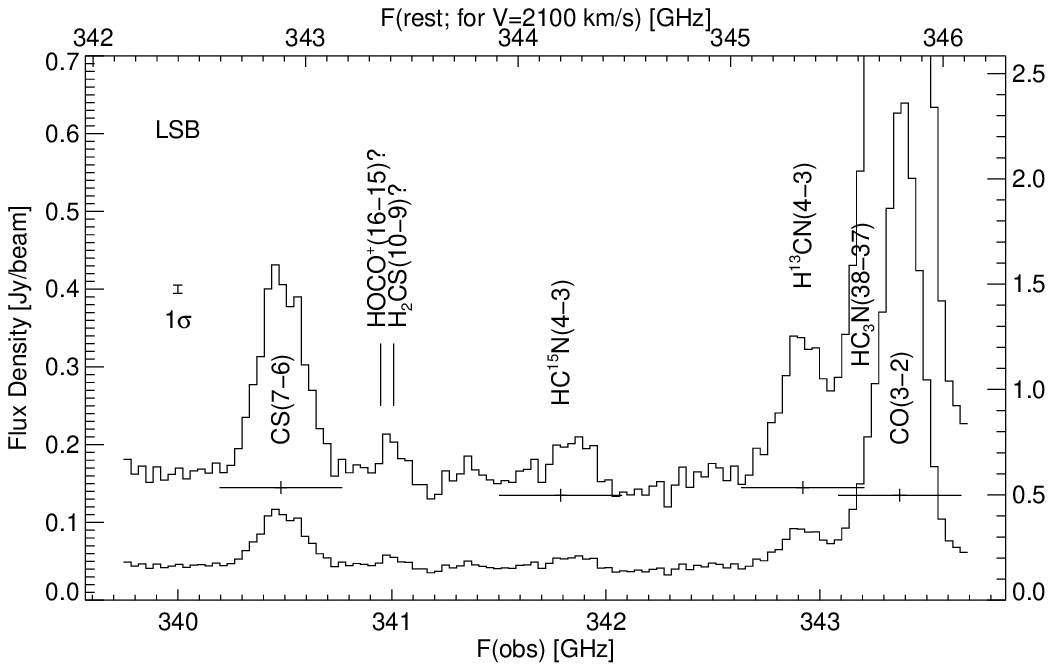}{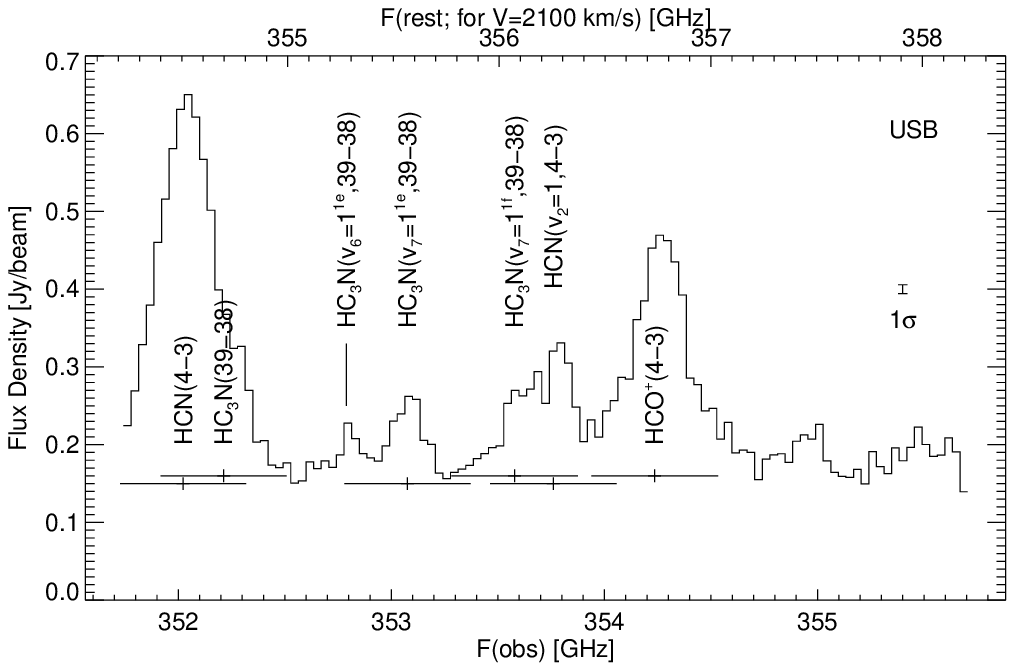} 
\caption{ \label{f.345SubExtspec}
NGC 4418 spectra in the central \about1\arcsec\ (data ID 09 and 10 in Table \ref{t.data}).
No continuum subtraction has been made. 
The same spectrum is plotted twice for the LSB data to show weak features (left scale)
and  the full spectrum (right scale).  
Horizontal bars under strong lines indicate the velocity ranges of 1850--2350 \kms, which are excluded
when making continuum data.   
The central tick of each bar is at 2100 \kms.
}
\end{figure}

Figures \ref{f.345VEXspec} and \ref{f.345SubExtspec} show
spectra\footnote{Throughout this paper, velocities are with respect to the local standard of rest (LSR) and
are defined with the radio convention, i.e., 
$v_{\rm radio} = c (1 -(f_{\rm obs}/f_{\rm rest}))$ where $f$ denotes frequency and $c$ the speed of light.
The conversion to velocity in the optical definition is $v_{\rm opt}=cz= v_{\rm radio} + 15\; \kms$ for NGC 4418. } 
made from the VEX data alone and from the EXT and SC data,  respectively.
The SC$+$EXT data are two times more sensitive than the SC data alone reported in \citest{Sakamoto10}. 

Two lines among about ten in the spectra  are new detections in this galaxy. 
Both are from vibrationally excited \HCthreeN.
Namely, \HCthreeN($v_7$=$1^{1f}$,J=39--38) is at \frest=356.072 GHz and has its 
upper energy level at $E_{\rm u} = 663 $ K.
\HCthreeN($v_6$=$1^{1e}$,J=39--38) is at \frest=355.278 GHz and has $E_{\rm u} = 1059 $ K.  
The doublet counterpart of the former line ($v_7$=$1^{1e}$,J=39--38) was already identified 
at \frest=355.566 GHz in \citest{Sakamoto10}.
Lower $J$ transitions of the latter line were detected in \citet{Costagliola10} and \citest{Costagliola12}.
Its doublet counterpart ($v_6$=$1^{1f}$,J=39--38) at \frest=355.557 GHz is blended with 
the \HCthreeN($v_7$=$1^{1e}$,J=39--38) mentioned above.
Although unidentified, there may be more lines, e.g., at \frest\ \about\ 343.4 and 357.5 GHz;
species having a line at the former frequency include H$_2$CS and HOCO$^{+}$.

The 350 GHz continuum in each sideband was made by averaging the following channels. 
For the VEX data we used channels outside of 1900--2300 \kms\ for
CO(3--2), HCN(4--3), \HthirteenCN(4--3), \HCfifteenN(4--3), \HCNv, \HCOplus(4--3), CS(7--6),
and \HCthreeN(39--38)  and (38--37). 
For the EXT and SC data we excluded channels having velocities of 1850--2350 \kms\ for the same lines\footnote{
The velocity range used here is the same as the one used to analyze the SC data
in \citest{Sakamoto10} although the note in their Table 1 described it otherwise.}
plus \HCthreeN($v_7$=$1^{1f}$,J=39--38) and ($v_7$=$1^{1e}$,J=39--38), 
channels between CO(3--2) and its nearest band edge, and the same for HCN(4--3).
The 660 GHz continuum was averaged over the full bandwidth and both sidebands.

%%%%%%%%%%%%%%%%%%%%%%%%%
\subsection{Concentration of Gas and Dust in the Nucleus}
\label{s.gas_dust_concentration}
Submillimeter emission is detected only from the central kpc (\about6\asec)
with different spatial extents in various emission components. 
We show this below using maps, aperture photometry, and visibility fitting.

%%%%%%%%%%
\subsubsection{Maps}
\begin{figure}[t]
\epsscale{1.0}
\plotone{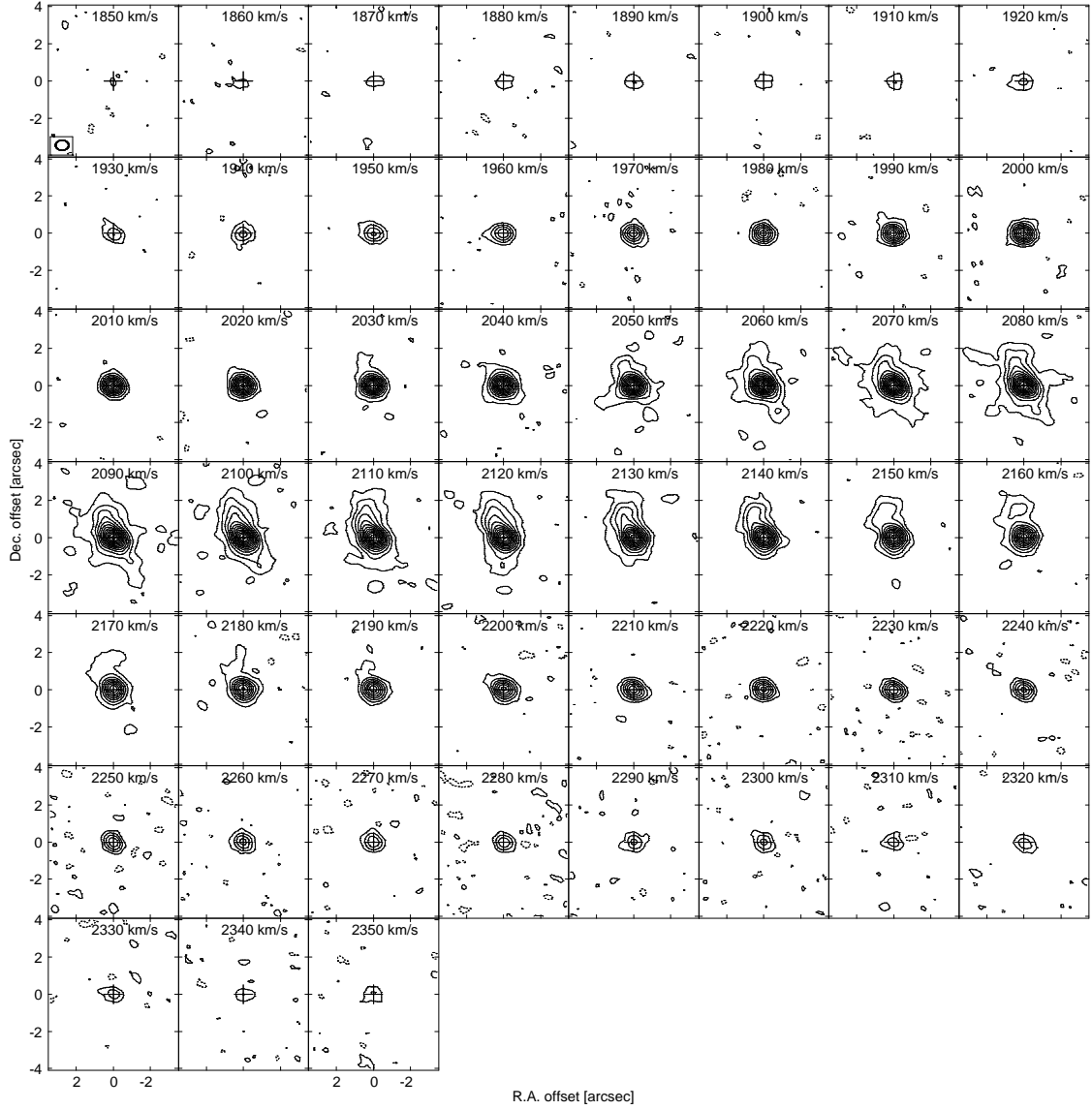} 
\caption{ \label{f.COchan_SubExtVex}
CO(3--2) channel maps made with the data from three array configurations.
The synthesized beam of 0\farcs69$\times$0\farcs55 (FWHM) is shown in the top-left panel.
The contours start from $\pm$35 mJy \perbeam\ = 0.95 K = 2.5 $\sigma$ and
the $n$th contours are at $35 n^{1.5}$ mJy \perbeam.
Negative contours are dashed.
The maximum intensity is 1.7 Jy \perbeam\ = 46 K at 2090 \kms.
The plus sign in each panel and the origin of the offset coordinates are at
the position of the 860 \micron\ continuum peak given in Table \ref{t.4418param}.
}
\end{figure}

%%% 
Figure \ref{f.COchan_SubExtVex} shows CO(3--2) channel maps to display faint extended emission at 0\farcs6 resolution. 
The data cube was made by combining all our data.
We detected here $797 \pm 80$ Jy \kms\ or $84\pm18$ \% of the single-dish CO(3--2) flux observed 
in the central 14\arcsec\ of the galaxy by \citet{Yao03}. 
Despite the high recovery rate, the CO(3--2) emission was detected only 
in the central 6\arcsec\ (1 kpc) of the galaxy, showing that the extent of the CO(3--2) emission is
intrinsically smaller than our 34\arcsec\ (5.6 kpc) field of view. 
The CO emission is not only compact but also strongly peaked toward the nucleus within the central kpc. 
In this data set 
more than half (58 \%) of the CO(3--2) flux in the central kpc is in the central 165 pc (1\asec, FWHM).

\begin{figure}[t]
\plotone{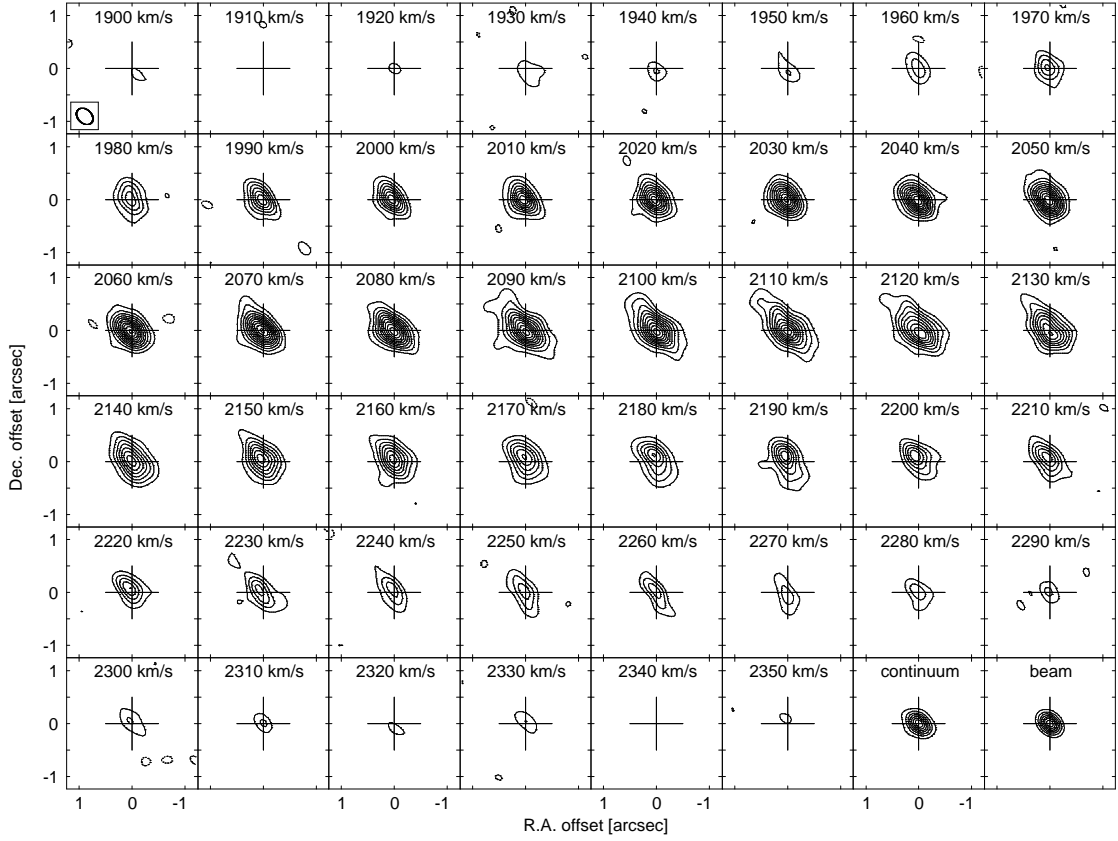} 
\caption{ \label{f.VEXCOchan}
CO(3--2) channel maps at the center of NGC 4418.
Only the data from the VEX configuration are used.
Contours are at \minus3, 3, 6, 9, 12, $\dots$, 33$\sigma$, where $3\sigma = 65$ mJy \perbeam\ = 7.3 K.
Negative contours are dashed.
The maximum intensity is 730 mJy \perbeam\ = 82 K at 2060 and 2070 \kms.
The plus sign in each panel and the origin of the offset coordinates are at
the position of the 860 \micron\ continuum peak given in Table \ref{t.4418param}.
The synthesized beam of 0\farcs35$\times$0\farcs26 (FWHM) is shown in the top-left panel.
The last two panels show LSB continuum and the point spread function (i.e., CLEAN beam), 
respectively, in contour steps of 12.5\% of each peak.
}
\end{figure}

\begin{figure}[t]
\plotone{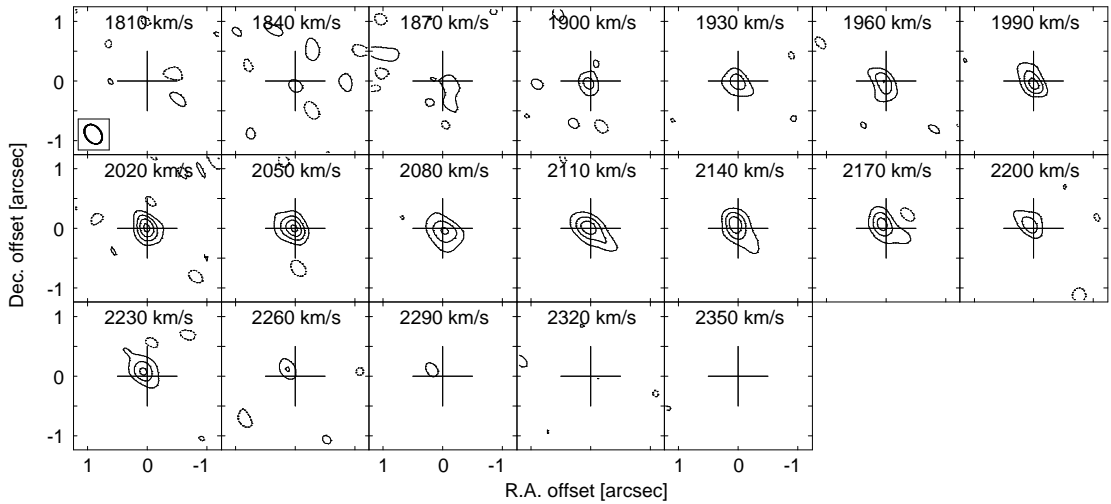} 
\caption{ \label{f.VEXHCNchan}
HCN(4--3) channel maps at the center of NGC 4418.
Only the data from the VEX configuration are used.
Contours are at (\minus1,1,2,3,4) $\times$ (50 mJy \perbeam = 5.4 K = 2.5$\sigma$).
Negative contours are dashed.
The maximum intensity is 23 K at 2020 and 2050 \kms.
The plus sign in each panel and the origin of the offset coordinates are at
the position of the 860 \micron\ continuum peak given in Table \ref{t.4418param}.
The synthesized beam of 0\farcs36$\times$0\farcs26 (FWHM) is shown in the top-left panel.
}
\end{figure}

Figures \ref{f.VEXCOchan} and \ref{f.VEXHCNchan} respectively are CO(3--2) and HCN(4--3) channel maps
at  \about0\farcs3 (50 pc) resolutions
showing that the line emission is strongly peaked at the nucleus at subarcsecond ($\lesssim$100 pc) scales.
These data cubes were made from the VEX data alone. 
This CO(3--2) data cube contains about 50\% of the CO flux detected in our 0\farcs6 resolution data.

Our continuum and line moment maps 
are shown in  Figs. \ref{f.SEVCOmom} and \ref{f.VEXmaps}. 
The spatial resolutions are about 0\farcs6 and 0\farcs3 for 350 GHz data
and 0\farcs2 for 660 GHz continuum. 
In particular, Fig. \ref{f.VEXmaps} (k) and (l) show our detection of a compact continuum source 
at the galactic center.
The continuum position is listed in Table \ref{t.4418param} and agrees to \about0\farcs3 with
the optical and radio positions of the nucleus;  see the caption of Table \ref{t.4418param} for the optical
position and \citet{Condon90} for the radio position.

%%%%%%%%%%%
\subsubsection{Aperture Photometry}
\label{s.app_photo}
The degree of central concentration, or compactness, is compared among various lines  and continuum
at 350 GHz
in Table \ref{t.param} using the ratio of flux (or flux density) between the central 0\farcs5 (80 pc)
and 6\arcsec\ (1 kpc). 
The former photometry is made using our data from the VEX configuration
and the latter data are from \citest{Sakamoto10} that used the SC configuration.  
The ratios for various emission components indicate their relative compactness.
The relatively extended nature of CO(3--2) emission is evident from its low ratio. 
Only one third of the CO(3--2) emission is from the central 0\farcs5. 
Lines other than CO have higher flux ratios, and hence are more concentrated toward the nucleus.
The ratio for 860 \micron\ continuum, $0.95\pm0.07$, is the highest    
among the ratios with S/N $>$ 3 in Table \ref{t.param}.
Thus 860 \micron\ continuum emission is more compact than both the CO and non-CO high excitation lines.
At the same time, the continuum emission in the VEX data is slightly more extended 
than the 0\farcs3 synthesized beam as seen in the last two panels of Fig. \ref{f.VEXCOchan}.
The detected 450 \micron\ continuum also looks compact but it has only a minor fraction of single-dish
measurement.

%%%%%%
\subsubsection{Visibility Fitting : 860 \micron\ Continuum}
\label{s.visfit_cont}
\begin{figure}[t]
\begin{center}
\epsscale{0.35}
\plotone{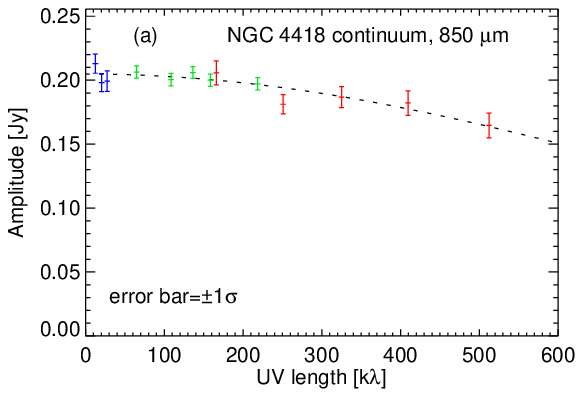} 
\epsscale{0.35}
\plotone{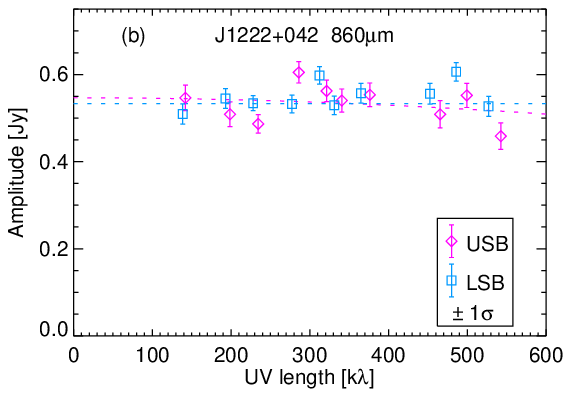} 
\epsscale{0.35}
\plotone{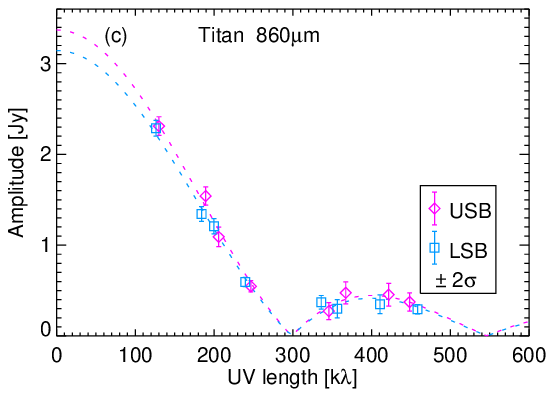} 
\vspace{1mm}
\epsscale{0.35}
\plotone{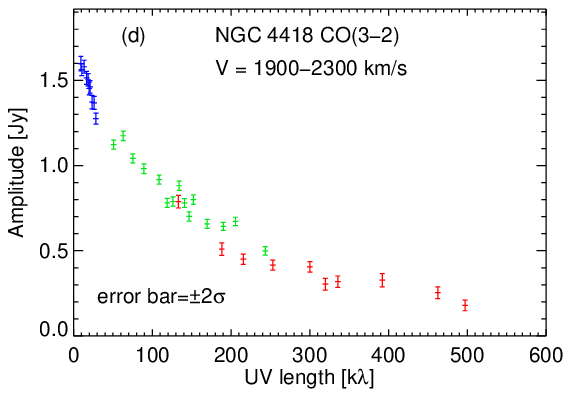} 
\epsscale{0.35}
\plotone{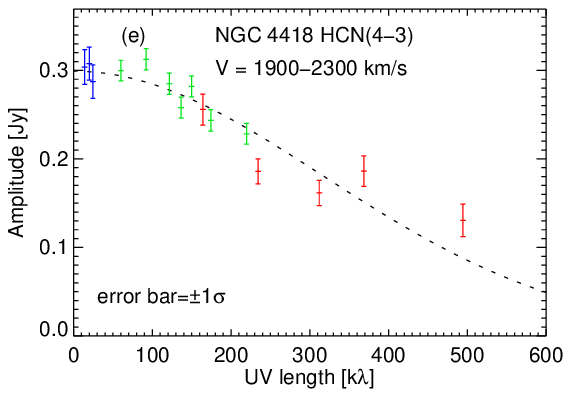} 
\epsscale{0.35}
\plotone{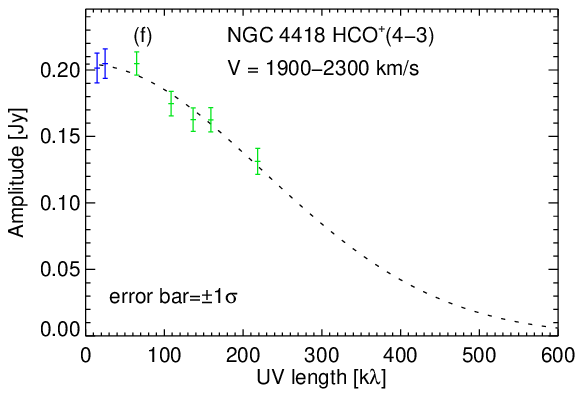} 
\end{center}
\caption{ \label{f.visfit}
Visibility fitting results.
(a)--(c): Continuum data of NGC 4418, J1222\plus042 (a test point source), and Titan (a source with
known size). The latter two show 2010 VEX data alone.
Dotted lines are Gaussian fits for (a) and (b) and a fit with a circular disk of known diameter for (c).
(d)--(e): Line data of NGC 4418 integrated over 1900--2300 \kms\ after continuum subtraction.
Data shown in blue, green, and red are from the SC, EXT, and VEX configurations of the SMA, respectively.
Dotted lines are models fitted to the data. 
See Table \ref{t.visfit} for the fit parameters.
The CO data could not be reasonably fit with a single Gaussian.
}
\end{figure}

Figure \ref{f.visfit} (a)--(c) show fits to the continuum visibilities of NGC 4418,
a test point source J1222\plus042, and Titan whose subarcsecond size is known.
The fit results are listed in Table \ref{t.visfit}.
For NGC 4418, data from the three array configurations are combined. 
This cannot be done for the calibrators because their flux densities and the apparent diameter of Titan change with time.
Since the NGC 4418 data were taken in three different nights the three data sets should have independent
flux calibration errors, which are typically about 10--20\% for the SMA 350 GHz observations.
However, we did not add the flux scaling error in the plot and in our fitting because doing so would
make the reduced $\chi^2$ well below unity. 
It is most likely that we achieved better flux calibration than usual,
presumably because we had accurate pointing from repeated pointing on the bright quasar 3C273 next to NGC 4418, 
our primary flux calibrator Titan was also close to 3C273 and NGC 4418, 
and atmospheric absorption was low (225 GHz zenith opacity $\leq 0.05$ for all the three nights). 
The size of 850 \micron\ continuum emission of NGC 4418 obtained from our data is 0\farcs10 (FWHM). 
We will use this for the source modeling in \S \ref{s.param_cont_core}.

We verified our visibility fitting using the quasar J1222\plus042 and Titan.
They were gain calibrated using 3C273 in the same way as NGC 4418 was calibrated. 
Our fit of J1222\plus042 using a Gaussian shows that blurring of the point source 
was less than 50 milli arcsec (mas) in our observations.
The diameter of Titan at 860 \micron\ is known to be 5230 km or 0\farcs8425 for the distance of 8.5588 AU.
This size includes the 40 km altitude of the tropopause, around which is the sub-mm photosphere of Titan.
The distance at the time of our 2010 VEX observations is from JPL-Horizons.
Fig. \ref{f.visfit} (c) shows that our data match the expected visibility curve very well with only flux scaling.
When the data were fit for both the source size and amplitude scaling, 
we obtained the size of the sub-arcsecond source with an
accuracy better than 3\% or to 20 mas.

The flux density of Titan sets the absolute amplitude scaling of all the visibility fitting.
We adopt the continuum brightness temperature of 77 K as the Planck temperature that gives the total flux density
of Titan for the solid-surface diameter (5150 km) of the satellite (M. Gurwell, private communication).
Our Titan data do not contain CO(3--2) and HCN(4--3) lines, both of which are known to be bright. 
Still, our imperfect knowledge of the atmosphere of Titan may introduce a common amplitude error to our data 
even though the relative calibration using Titan is precise and consistent among the three data sets. 
We therefore assign 5\% to the overall flux scaling uncertainty of our NGC 4418 fit, even though
the formal error in the fitting was about 1\%.
In addition to this, there is likely 
contamination of line emission to the `line-free' channels  used for the USB continuum of the galaxy.
We observed a sign of that in our SC configuration data 
in the form of an uncomfortably large spectral index $\alpha=6.25 \pm 0.84$ ($S_\nu \propto \nu^\alpha$)
between the two sidebands \citesp{Sakamoto10}. 
A similar sign was seen in our 2010 VEX data too.
Since we used the USB continuum in the multi-configuration fitting as LSB data were noisier,
the line contamination is likely in the fitted amplitude.
Its effect on the luminosity estimate is addressed in \S \ref{s.Tb_Lbol}.

%%%%%%
\subsubsection{Visibility Fitting : Lines}
\label{s.visfit_line}
\begin{figure}[t]
\epsscale{0.6}
\plottwo{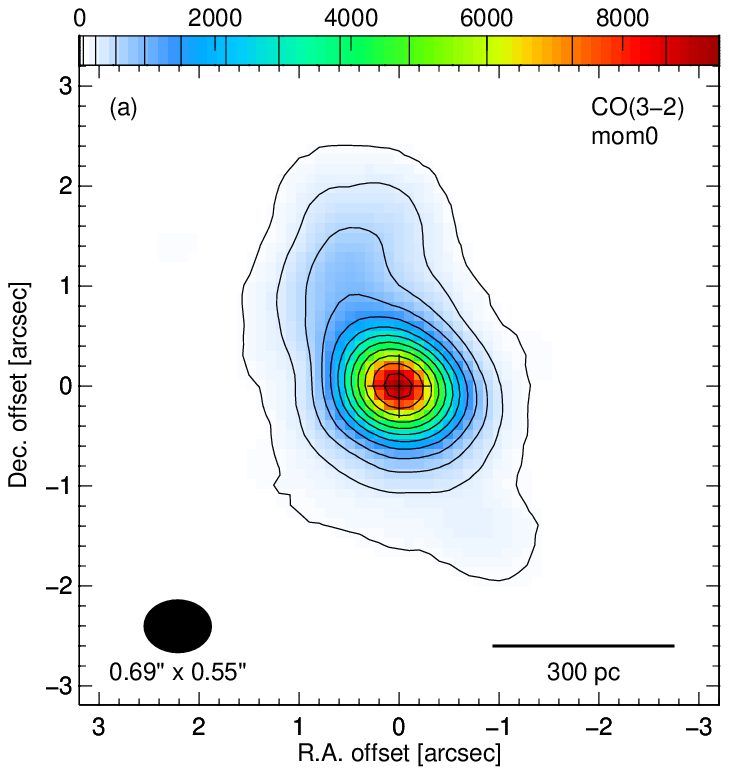}{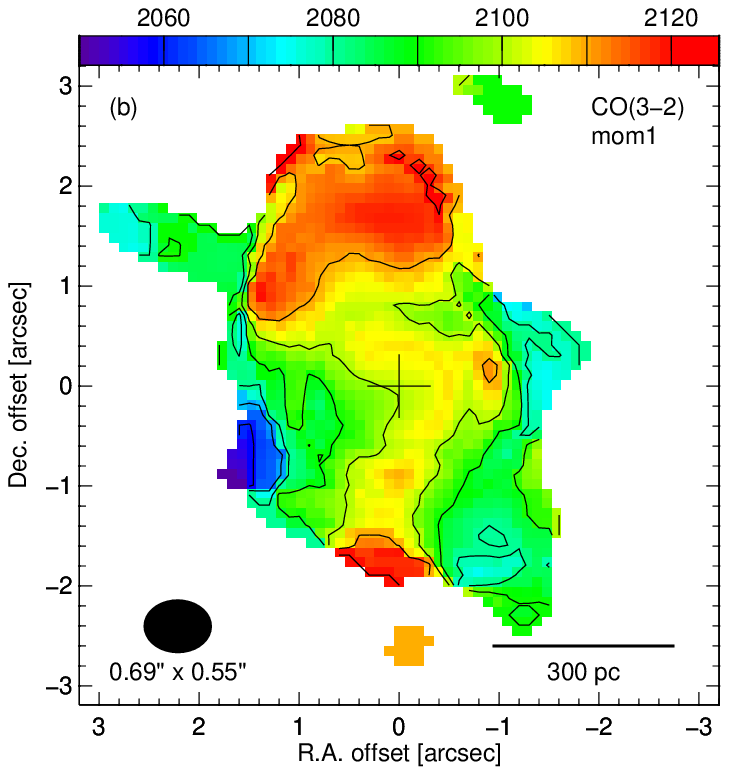}  
\epsscale{1.0}
\caption{ \label{f.SEVCOmom}
CO(3--2) maps made using three array configurations.
(a) Integrated intensity. The $n$th contour is at $60 n^2$ K \kms. The peak intensity is $9.4\times10^3$ K \kms.
(b) Intensity-weighted mean velocity. Contour interval is 10 \kms.
The plus sign in each panel and the origin of the offset coordinates are at
the position of the 860 \micron\ continuum peak given in Table \ref{t.4418param}.
The black ellipses show the FWHM of the synthesized beam.
}
\end{figure}

Fig. \ref{f.visfit} also shows visibility amplitudes of line emission against baseline lengths.
Line visibilities were averaged over 1900--2300 \kms\ after continuum subtraction, and then binned
according to the baseline length. 
It is evident in the figure that the line visibility amplitudes decline faster than the continuum amplitudes. 
In other words, line emission is more extended than continuum.
Comparison of CO, HCN, and \HCOplus\ data tells us that the CO emission is more extended
than the latter two, confirming our observation using the aperture photometry.
Moreover, the CO visibility plot shows that the data cannot be fit with a single Gaussian in the \uv\ domain.
Hence the spatial distribution of the CO(3--2) emission is not a Gaussian.
The line emission distribution has at least a compact subarcsecond peak toward the nucleus 
(Fig. \ref{f.VEXmaps}) and an extended component with the total extent of \about 5\arcsec\ (Fig. \ref{f.SEVCOmom}).
The extended component is non-axisymmetric around the nucleus (Fig. \ref{f.SEVCOmom}).

%%%%%%%%%%%%%%%%%%%%%%%%%%%%%%%%%%%
\subsection{Parameters of the Continuum Core}
\label{s.param_cont_core}

%%%%%%
\subsubsection{Spectral Energy Distribution, Continuum Opacity}
\label{s.SED}
\begin{figure}[t]
\epsscale{0.7}
\plotone{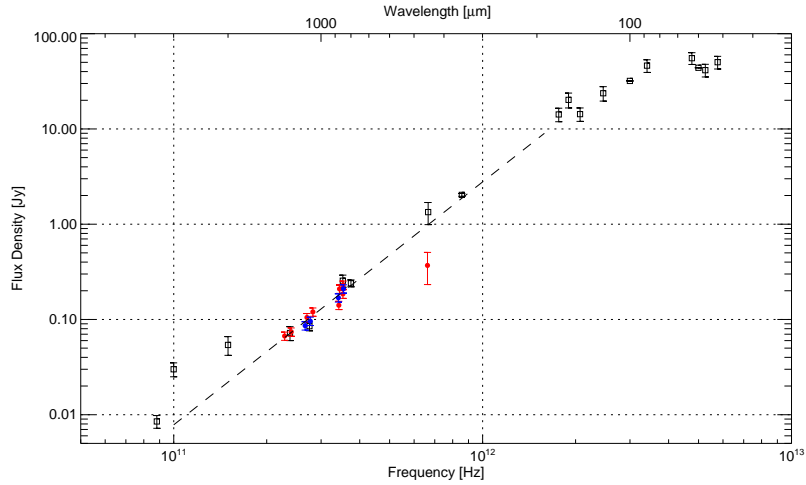} 
\caption{ \label{f.sed}
Spectral energy distribution of NGC 4418 in its Rayleigh-Jeans part.
Filled data points are from SMA observations; 
blue ones are from the short baseline observations in \citet{Sakamoto10} 
and red ones are from long baseline observations. 
Error bars are $\pm 1 \sigma$.
The dashed line is a power law fit to the SMA data between 200 and 400 GHz.
It has a spectral index of $\alpha=2.55 \pm 0.18$ (for $S_\nu \propto \nu^\alpha$) 
and is extrapolated to a wider frequency range to guide eyes.
Non-SMA data are from the literature 
\citep{Brauher08,Sanders03,Yang07,Roche93,Dunne00,Imanishi04}. 
No subtraction for line contamination is made for the bolometer data between 100 GHz and 1 THz.
}
\end{figure}

The continuum SED has a power-law slope
of $\alpha=2.55 \pm 0.18$ ($S_\nu \propto \nu^\alpha$) between 1.3 mm and 0.85 mm.
This is measured from the SMA data in Figure \ref{f.sed}, 
namely from  this work, \citest{Sakamoto10}, 
and the 1.3 and 1.1 mm observations used in the companion paper \citesp{Costagliola12}.
Since the SMA continuum data were taken from line-free spectrometer channels in each dataset,
none of the SMA data is contaminated by strong emission lines such as CO 
unlike wide-band bolometer data.
The spectral slope that we obtained is shallower (i.e., $\alpha$ is smaller) than those of other local IR luminous galaxies
but is consistent with the results of the FIR-to-mm SED surveys of \citet{Lisenfeld00} and \citet{Yang07},
in each of which NGC 4418 has the shallowest spectral slope in about 15 galaxies.

Two simple models for the spectral index suggest that the $\alpha$ measured
at the nucleus is due to the dust opacity $\tau \sim 1$ at $\lambda \sim 1 $ mm.
In the first model of a simple one-zone slab, the
spectral slope should be $\alpha = 2 + \beta \tau (e^\tau -1)^{-1}$
where $\beta$ is the index of power-law frequency dependence of the dust opacity
and $\tau$ is the slab opacity at the wavelength of the $\alpha$ measurement.
Here we used the Rayleigh-Jeans approximation because the brightness temperature of the
0\farcs1 nucleus exceeds 100 K as we will see next (\S \ref{s.Tb_Lbol}).
With optically thin emission, this model gives the familiar form of $\alpha = 2 + \beta$.
The low $\alpha$ observed in NGC 4418 means low $\beta$ or 
a moderate opacity of dust emission, i.e., $\tau \sim 1$ at $\lambda \sim 1 $ mm.
The dust opacity about 1 is favored, as was concluded by
\citet{Lisenfeld00}, because most infrared luminous galaxies have $\beta \approx 1.5$--$2$
\citep{Lisenfeld00,Dunne01}. 
The observed $\alpha$ corresponds to $\tau=2.2 \pm 0.5$ and $1.8 \pm 0.5$ for $\beta=2$ and  $1.5$, respectively.
For the second model,
 we note that the SED slope does not become 2 at shorter wavelengths as expected for $\tau \gg 1$.
The shallower slope of the Planck function compared to the Rayleigh-Jeans formula must partly be the reason. 
Another likely reason is that the dust distribution is not just a 0\farcs1 slab of uniform opacity
but presumably has a low opacity halo as suggested by the larger extent of line emission than the 860 \micron\ continuum.
The spectral slope in the general multi-component case without mutual shielding
is the flux-weighted mean of the spectral indexes,
$\alpha = \sum S_i \alpha_i / \sum S_i$, where 
the $i$th component has the flux density $S_i$ and spectral index $\alpha_i$. 
For $\alpha_1=2$ and $\alpha_2 = 4$, 
the observed $\alpha$ corresponds to a fraction of the optically thick core of $73 \pm 9$ \%. 
Thus as long as we assume $\beta \approx 2$, we need an optically thick core ($\tau \gtrsim 1$) 
that dominates the spectral energy distribution at $\lambda \sim 1 $ mm.
In addition to the two models, the high opacity is also supported by the high brightness temperature 
that sets a lower limit of about 0.1 to the 860 \micron\ opacity from the constraint
that opacity-corrected brightness temperature cannot exceed dust sublimation temperature.

There are other factors that depend on frequency and potentially affect the spectral index. 
They include the core size defined with the $\tau=1$ surface,
the dust temperature at the photosphere, and absorption of the core emission by dust around it.
The last one is addressed in \S \ref{s.450cont}.

\subsubsection{Gas Column Density, Mean Density, and Mass} 
\label{s.NH_Mgass}
The dust continuum opacity $\tau_{\rm 860\; \mu m} \approx 1$ of the core translates to 
a hydrogen column density $ \NH \sim 10^{25.7}$ \persquarecm\  
for  the dust opacity to column density relation
$N({\rm H+\HH})/\tau_{\lambda} = 1.2 \times 10^{25} (\lambda/400\; \micron)^2$ H-atom \persquarecm\
\citep{Keene82,Hildebrand83}.
Since the core has a size of \about20 pc  (0\farcs1 in FWHM from \S \ref{s.visfit_line}; see the next subsection
for two models), we obtain $\nHH \sim 5\times10^{5}$ \percubiccm\
for the mean gas density (i.e., number density of hydrogen molecules) 
and 
$\Mmol(r\leq 10\;{\rm pc}) \sim 1 \times 10^{8}$ \Msol\ for the total gas mass in the core  
assuming that the gas is mostly molecular there.
The total mass depends on the density distribution. 
The estimate above is for a uniform density and is an upper limit 
if the gas density increases toward the center.

%%%%%%
\subsubsection{Brightness Temperature and Luminosity} 
\label{s.Tb_Lbol}
Table \ref{t.lbol} lists the deconvolved (peak) brightness temperature 
and the bolometric luminosity of the nucleus. 
The luminosity is derived from the 860 \micron\ core size and the temperature.
Here we use two different models to estimate the brightness temperature and the bolometric luminosity of the
nucleus as we did in \citest{Sakamoto08} for Arp 220.
One is the circular Gaussian we used for the visibility fit above and the other is
a circular disk with uniform brightness. 
The Gaussian model describes dust that is not fully opaque across the nucleus and 
that may also have a radial temperature gradient.
The circular disk model corresponds to a sphere of uniform surface brightness projected on the sky.
Both models can fit the same data well when the
visibilities are sampled only within the half-maximum 
baseline length of the visibility function. 
The diameter of the model disk is 1.6 times the
FWHM of the Gaussian in such a case \citesp{Sakamoto08}.
For the luminosity calculation, it is assumed that the 860 \micron\ emission is optically thick
in the uniform disk/sphere model. 
Since the deconvolved \Tb\ was used without any opacity correction (i.e., not divided by $1-e^{-\tau}$), 
the derived temperature and luminosity are lower limits. 
We calculated the luminosity of the Gaussian model by using its FWHM size and its peak brightness temperature.
This crudely approximates the effect mentioned earlier that the size of the photosphere should increase 
toward higher frequencies as dust opacity increases with frequency.

The derived luminosity of the nucleus is about $1\times10^{11}$ \Lsol, which is comparable
to the total luminosity of the galaxy ($10^{11.1}$ \Lsol, Table \ref{t.4418param}).
We obtained nuclear luminosities 
$\log (\Lbol/\Lsol) =  11.3 \pm 0.3$ and $10.8 \pm 0.3$ for the two models.
Luminosity surface density and volume density are also calculated for each model and listed
in Table \ref{t.lbol} for the central \about20 pc.
The errors here and in Table \ref{t.lbol} are random errors 
due to the 5\% flux error and the 11\% size uncertainty 
and do not include systematic error due to the models.

There are two notable sources of systematic errors in the derived parameters in addition to 
the assumptions on the source shape and on the 860 \micron\ opacity.
(The underestimating effect of the latter is minor as noted in \citest{Sakamoto08}.) 
One is the absorption of the core emission by the surrounding dust and gas. 
It arises because the \about0\farcs1 (\about20 pc) continuum core is embedded 
in an order of magnitude larger envelope as we saw in \S \ref{s.gas_dust_concentration}.
The continuum emission from the core must suffer from extinction to some extent by the envelope 
through continuum self-absorption and line absorption by various molecules.
Our continuum brightness temperature derived without the extinction correction
is therefore a lower limit of the true brightness temperature at the surface of the \about0\farcs1 core.
The correction for it would increase the core luminosity.
We will analyze this submillimeter extinction a little more in \S \ref{s.450cont} and find
the extinction to be about 5\% at 860 \micron.
The other source of systematic error is the possible contamination of the 860 \micron\ continuum
by molecular lines. 
This effect can be estimated to be about 10\% assuming that the spectral index between
lower and upper sidebands obtained in the SC configuration, $6.25\pm0.84$ (\S \ref{s.visfit_cont}),
is due to line contamination to the USB data and that the true continuum spectral index is
$2.55\pm0.18$ (\S \ref {s.SED}).
The continuum brightness temperature of the core would be lower without the contamination.
However, this does not necessarily mean that the bolometric luminosity of the core is 
overestimated by this effect. 
Thermal lines can significantly add flux density to the 860 \micron\ continuum
only when the line and continuum opacities satisfy 
$\tau_{\rm line} > \tau_{\rm cont}$ and $\tau_{\rm cont} \lesssim 1$, because no thermal line will be seen
from a blackbody.
Since we use the brightness temperature of the continuum core for our luminosity estimate
assuming that the continuum is optically thick, the line emission, if it is from the 0\farcs1 core itself, 
would (partly) compensate for the decrement
of the observed continuum brightness temperature due to $\tau_{\rm cont}$ that is not $\gg 1$.

In addition to these two error sources, 
our visibility fits are insensitive to a point-like source 
with a small flux contribution.
This is due to our \uv\ coverage limit.
For example, a 20 mJy source with a size of 20 mas (3 pc)
can be undetected in our visibility fit (Fig. \ref{f.visfit} a) and would have
a temperature of 700 K and luminosity of $10^{12}$ \Lsol.  
The total luminosity of the galaxy limits the presence of such compact sources.

%%%%%%%%%%%%%%%%%%%%%%%%%
\subsubsection{450 \micron\ Continuum}
\label{s.450cont}
We detected 450 \micron\ continuum at 5.8$\sigma$ with the resolution of 0\farcs23 $\times$ 0\farcs15
at the position of the 860 \micron\ continuum peak (Fig. \ref{f.VEXmaps} (l)).
The 450 \micron\  nucleus in the map appears unresolved but this is largely due to the low signal-to-noise (S/N) ratio.
(Image-domain fitting gives a nominal deconvolved size of 0\farcs10$\pm$0\farcs04 in the major axis
while the source is unresolved in the minor axis.)
The 450 \micron\  detection of the subarcsec-scale nucleus is at 10$\sigma$ and hence even more secure when 
the data are convolved to 0\farcs5 resolution (Table \ref{t.param}).
(The S/N increases because the convolution lowers the weight of noisy long baselines.)  
The peak and total flux densities in both resolutions are about 370 mJy and mJy \perbeam, respectively.

\begin{figure}[t]
\epsscale{0.4}
\plotone{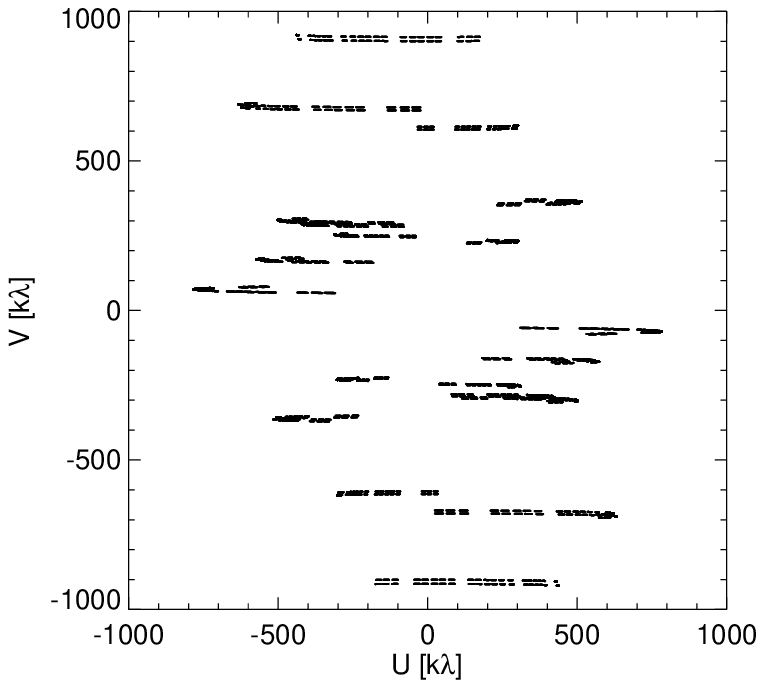} \\   
\epsscale{0.4}
\plotone{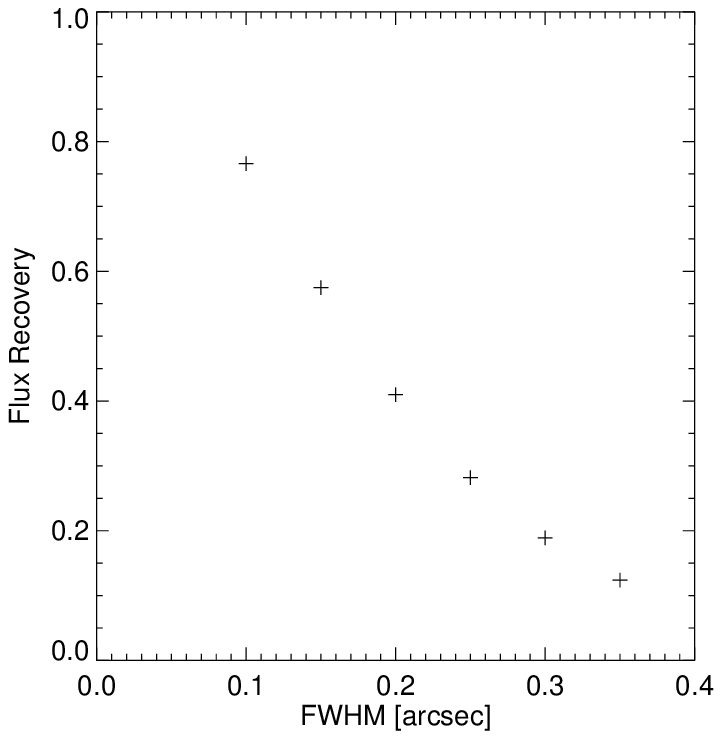}    
\caption{ \label{f.FluxRecv660}
(Top) Sampling of our 660 GHz observations in the \uv\ plane.
(Bottom) Fraction of flux recovered in the simulations of our observations for
Gaussian sources with various sizes.
}
\end{figure}

The 450 \micron\ continuum flux density that we detected is significantly lower than 
that of the central 3 kpc of the galaxy  (see Fig. \ref{f.sed}). 
The ratio is  $0.31 \pm 0.15$ between the SMA and single-dish measurements.
The latter was made by \citet{Roche93} with the UKT14 bolometer,
the filter of which had the effective frequency of 682 GHz (440 \micron) and the bandwidth of 84 GHz \citep{Duncan90}. 
We subtracted from the bolometer measurement the contribution of CO(6--5) line (\about0.1 Jy, calculated
from the CO(3--2) single-dish flux of \citet{Yao03} scaled by a factor of 4) and rescaled the continuum
to our observing frequency using the spectral index $\alpha=2.55$.
The single-dish flux density of the 450 \micron\ continuum estimated this way is $1.18 \pm 0.33$ Jy.
The low flux recovery ratio at 450 \micron\ suggests that there is a significant 450 \micron\ emission 
that is too extended for detection at our sensitivity and with the shortest baseline of  240 k$\lambda$ in the VEX configuration. 
We simulated the degree of flux recovery with our actual \uv\ coverage for
Gaussian sources of various sizes.
(We did not fit the visibilities unlike our 860 \micron\ data analysis because the 450 \micron\ data have lower S/N ratio.)
Fig. \ref{f.FluxRecv660} shows that the size of the 450 \micron\ continuum source would be $0\farcs24 \pm 0\farcs07$ (FWHM)
if the source were approximately a single Gaussian. 
The peak brightness temperature of the source would be $70 \pm 20$ K and
the bolometric luminosity of this source would be on the order of only $10^{10.2}$ \Lsol, 
which disagrees with our estimate at 860 \micron.

A two-component model with a core and an envelope agrees with the observations within the errors 
and is more reasonable in our view. 
The two components are introduced in view of the compact continuum core and the more extended CO
emission both observed at 860 \micron.
They are treated separately to simplify our modeling though they are probably not two discrete entities.
(More detailed multi-component models have been made by \citet{Gonzalez-Alfonso12} and \citest{Costagliola12}.
Our intention here is not to update their models but to apply a simple model to understand our sub-mm continuum observations.)
The core and envelope are assumed to have sizes of 0\farcs1 and 0\farcs5 (both in FWHM), respectively. 
Also the core is assumed to dominate the 860 \micron\ emission with 85\% of the total flux density. 
The extended envelope with the small remaining flux density has too low a surface brightness 
to be detected in our high-resolution map.
However, the amount of extended 860 \micron\ continuum in this model is implied by our CO(3--2) data.
Namely, if one third of the CO(3--2) emission is from the envelope region, then its flux of about 300 Jy \kms\
and  the typical CO(3--2) equivalent width of $1\times10^{4}$ \kms\ among infrared bright galaxies \citep{Seaquist04}
suggest a 860 \micron\ extended continuum of \about30 mJy, which is about 15\% of the total continuum flux density.
The analysis of spectral index in \S \ref{s.SED} also implied a small contribution to 860 \micron\ emission 
from an optically-thin and extended component. 
We assume the spectral index between 860 \micron\ and 450 \micron\ to be 2 for the optically thick core and
4 for the optically thin envelope (i.e., the model assumes $\beta=2$). 
In addition, since the core is embedded in the envelope, the former should be extinguished by the latter. 
The opacity of the envelope to cover the core is assumed to be \about1 at 200 \micron, which we took from 
the model of \citet{Gonzalez-Alfonso12} noting 
that it makes the envelope optically thin at 860 \micron\ ($\tau = 0.05$) as we assumed above.
At 450 \micron\ the core is extinguished by the envelope with $\tau=0.2$ and the envelope itself is resolved out in our VEX data.
Overall, this model predicts our flux recovery rate to be about 45\%, 
which agrees with our observed value of $31 \pm15$ \%.
The current data quality does not justify further parameter tuning and model refinement.

This model illustrates that two effects likely contribute to 
the lower flux recovery rate at 450 \micron\ than 860 \micron.
One is that the core emission is saturated while emission from the extended outer region is not.
The other is that the emission from the hot nucleus is extinguished more at shorter wavelengths 
by the colder circumnuclear dust.
These effects can also explain the 60 \micron\ peak of the SED of NGC 4418 despite
the energetically dominant core emitting at $> 100$ K at 860 \micron.
The same effects probably explain similar observations in Arp 220 where the two nuclei contribute much 
less to the total at 435 \micron\ than at 860 \micron\ \citep{Matsushita09}.

%%%%%%%%%%%%%%%%%%%%%%%%%
\subsection{Kinematics of Molecular Gas}
\label{s.kinematics}
\begin{figure}[b]
\begin{center}
\includegraphics[scale=0.9]{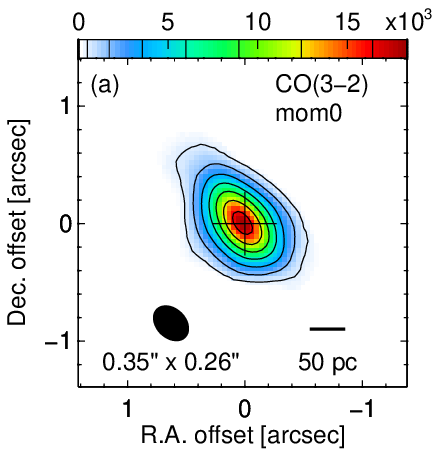} 
\includegraphics[scale=0.9]{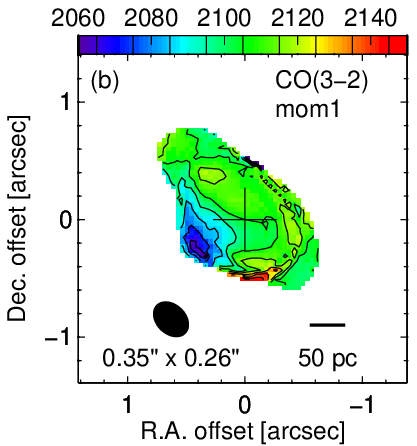} 
\includegraphics[scale=0.9]{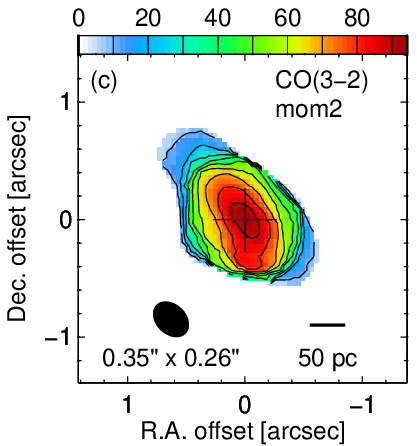} 
\includegraphics[scale=0.9]{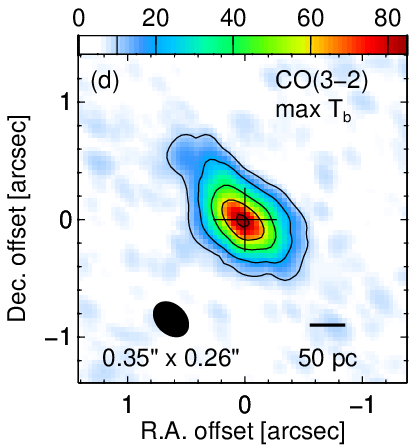} \\
\includegraphics[scale=0.9]{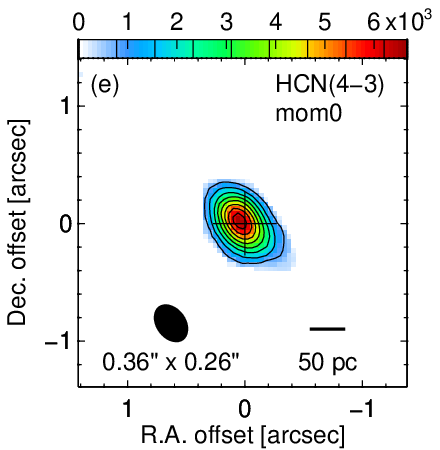} 
\includegraphics[scale=0.9]{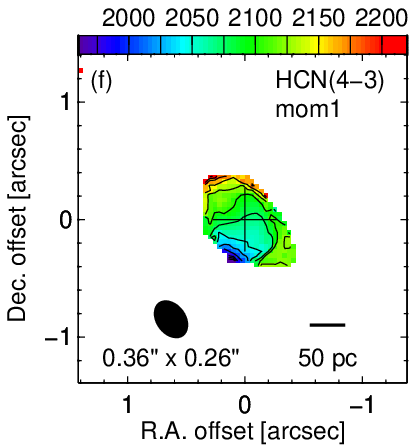} 
\includegraphics[scale=0.9]{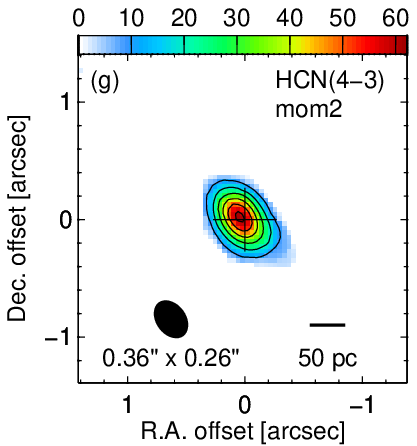} 
\includegraphics[scale=0.9]{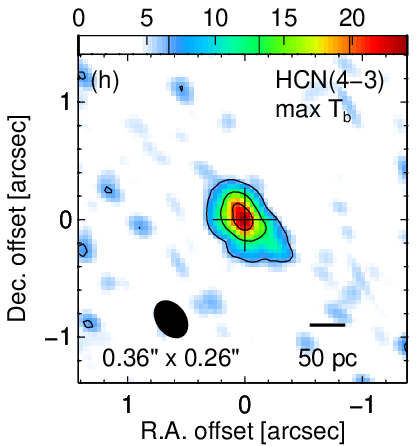} \\
\includegraphics[scale=0.9]{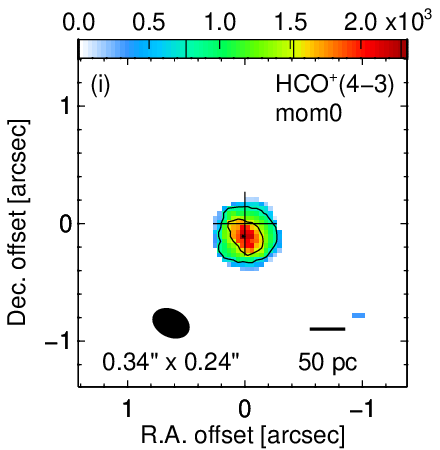} 
\includegraphics[scale=0.9]{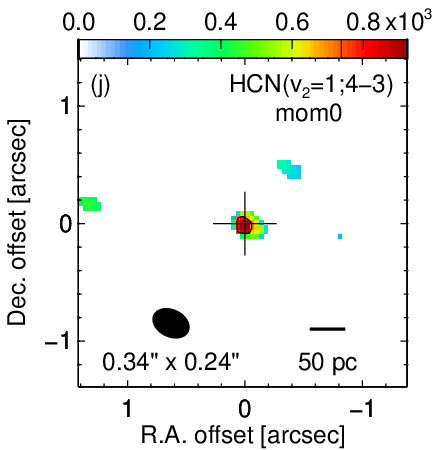} 
\includegraphics[scale=0.9]{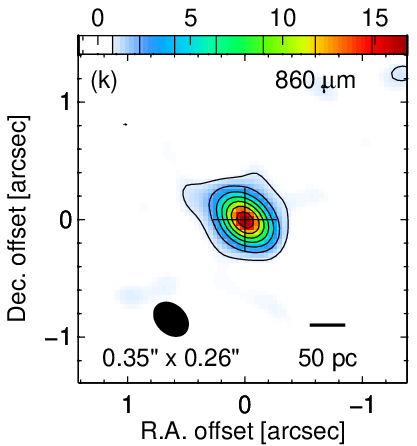} 
\includegraphics[scale=0.9]{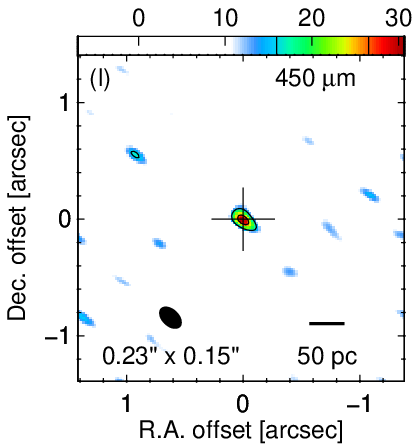}  
\end{center}
\vspace{-4mm}
\caption{ \label{f.VEXmaps}
Submillimeter line and continuum images at the center of NGC 4418;
(a)--(d) are CO(3--2), (e)--(h) are HCN(4--3), 
(i) is  \HCOplus(4--3),  (j) is \HCNv,  and (k) and (l) are 860 and 450 \micron\ (DSB) continuum, respectively.
Data from the VEX configuration are used.
The position of the 860 \micron\ continuum peak,  which is given in Table \ref{t.4418param},
is marked with the plus sign in each panel and is used as the origin of the offset coordinates.
The black ellipse in each panel shows the FWHM of the synthesized beam.
The images labeled with `mom0', `mom1', `mom2' show integrated intensity (i.e., 0-th moment) in K \kms, 
intensity-weighted mean velocity (1st moment) in \kms, and velocity dispersion (2nd moment) in \kms, 
respectively, and those labeled with `max \Tb' show peak brightness temperature in K.
The contour information is as follows.
(a) The $n$th contour is at $0.5 n^{1.8}\times 10^3$ K \kms. 
The peak value is $18\times 10^3$ K \kms.
(b) 10 \kms\ steps.  The contour crossing the continuum peak is at 2105 \kms.
(c) 10 \kms\ steps.
(d) 10, 20, 40, 60, and 80 K (1$\sigma$=2.4 K).  The peak is 82 K.
(e) $0.78 n$ $10^3$ K \kms. 
The peak is $6.7\times 10^3$ K \kms.
(f) 25 \kms\ steps. The contour crossing the continuum peak is 2075 \kms.
(g) 10 \kms\ steps.
(h) (1, 2, 3) $\times$ (6.6 K = 3 $\sigma$). The peak is 23 K.
(i) (1, 2, 3) $\times$ $0.76 \times 10^3$ K \kms. 
(j) The contour is at $0.74 \times 10^3$ K \kms. 
(k) $0.8 n^{1.5}$ K, starting at $3\sigma$. The intensity peak is 17 K.
(l)  Contours are at $(-3, 3, 5) \times (1\sigma$ = 5.3 K = 65 mJy \perbeam).
Peak = 31 K(R-J) = 374 mJy \perbeam\ = 5.8$\sigma$.
}
\end{figure}

\begin{figure}[b]
\epsscale{0.6}
\plottwo{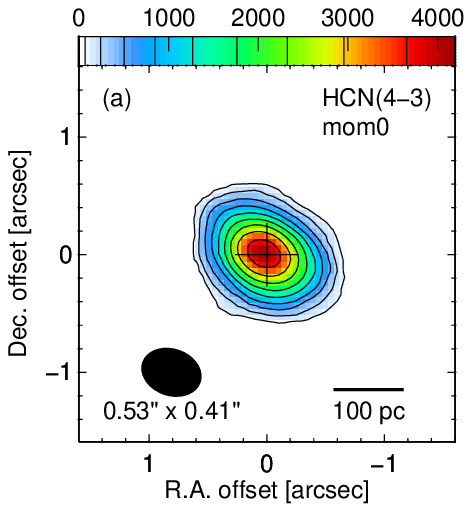}{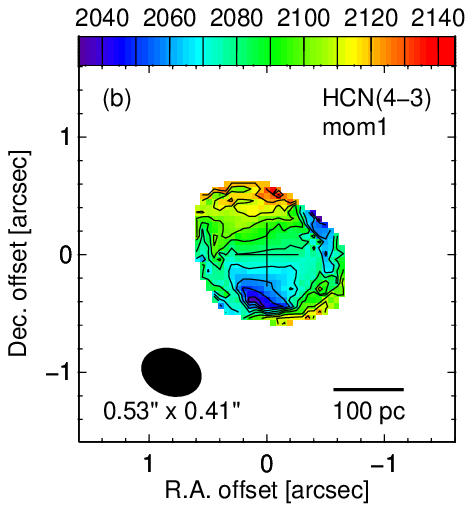} \\
\plottwo{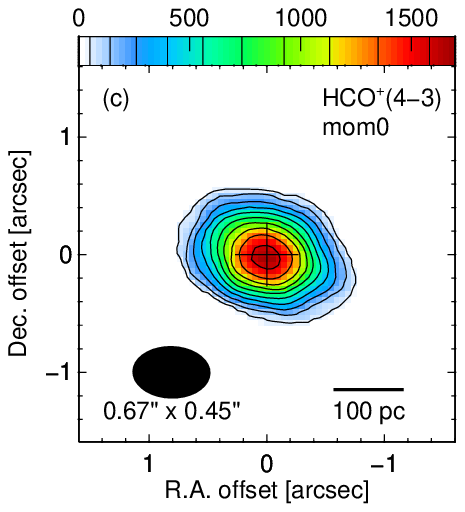}{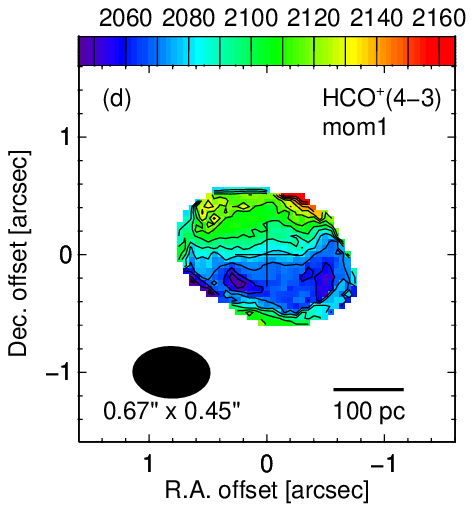} \\
\plottwo{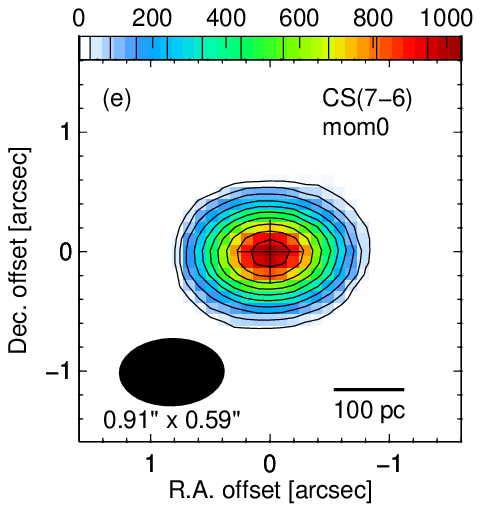}{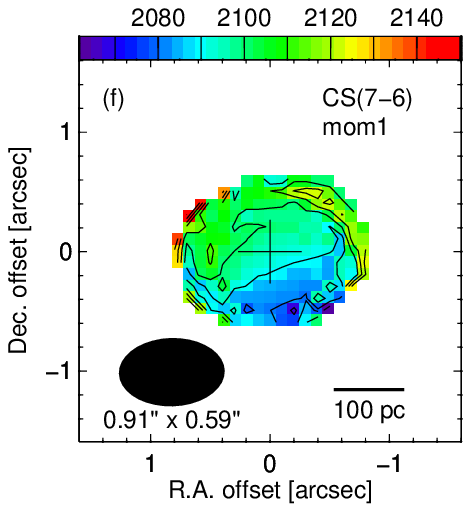} 
\caption{ \label{f.SEVHCNmom}
Integrated intensity and mean velocity maps of HCN(4--3), \HCOplus(4--3), and CS(7--6) emission.
Data from three array configurations are used except for CS that was observed in two. 
The $n$th contour is at $70 n^{1.8}$, $50 n^{1.5}$, and $30 n^{1.5}$ K \kms\ in (a), (c), and (e), respectively.
Velocity contours are in 10 \kms\ steps.
The position of the 860 \micron\ continuum peak,  whose coordinates are in Table \ref{t.4418param},
is marked with the plus sign in each panel and is used as the origin of the offset coordinates.
The black ellipses show the FWHM sizes of the synthesized beams.
}
\end{figure}

The HCN velocity map Fig.~\ref{f.VEXmaps}(f) shows in the central 0\farcs5 (80 pc)
a systematic gradient of mean velocity in the northeast--southwest direction,
which is close to the galaxy major axis (p.a.=60\arcdeg, Table \ref{t.4418param}).
The HCN channel maps in Fig. \ref{f.VEXHCNchan} show this velocity structure as
the shift of the emission peak across the continuum position from southwest (lower-right)
to northeast (upper-left) in channels between 1990 and 2200 \kms. 
In Figure \ref{f.SEVHCNmom}, moment maps of HCN, \HCOplus, and CS confirm the
presence of the gradient across the nucleus although the direction of the
velocity gradient is closer to the north--south direction in these lower resolution data.
This is also the case in the CO high-resolution data (Fig. \ref{f.VEXmaps} (b)) in the central 0\farcs5.

\begin{figure}[t]
\begin{center}
\includegraphics[scale=1.0]{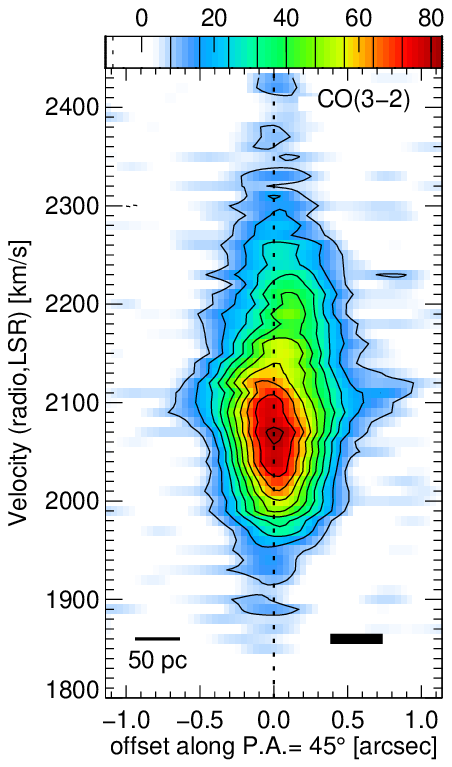} 
\includegraphics[scale=1.0]{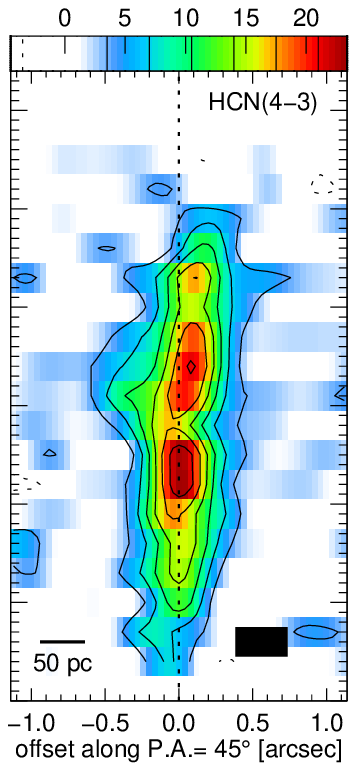} 
\includegraphics[scale=1.0]{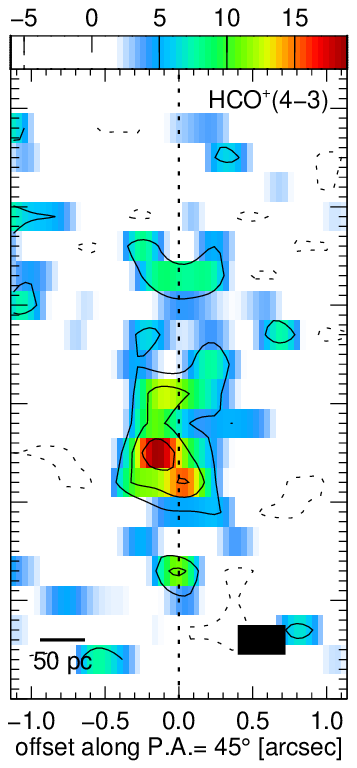} 
\end{center}
\caption{ \label{f.pv}
Position-velocity diagrams of molecular lines at the center of NGC 4418.
Each PV cut is through the continuum peak and along the position angle of 45\arcdeg.
The origin of the abscissa is the continuum peak.  
The black rectangles at the bottom-right show the spatial and velocity resolutions of the data.
Contour steps are 8 K (3.3$\sigma$) for CO, 3.5 K (1.6$\sigma$) for HCN, and 5 K (1.6$\sigma$)
for \HCOplus. 
The highest-velocity emission in the CO diagram is from \HthirteenCN, and the low-velocity
signal in the HCN diagram likely contains \HCthreeN(39--38).
}
\end{figure}

We attribute the central velocity gradient to rotation around the luminous core 
and estimate its dynamical mass to be 
$\Mdyn (r\leq {\rm 15\, pc}) \sim 2\times10^{8} \Msol$.
We estimated for this a velocity shift of about 400 \kms\ in 0\farcs2 across the nucleus
from the CO and HCN position-velocity diagrams in Fig. \ref{f.pv},
where almost the full velocity range of each line is observed at
the central \about0\farcs3.
The Keplerian dynamical mass above is for the 62\arcdeg\ inclination of the galaxy.
Because of the insufficient spatial resolution to obtain an accurate rotation curve,
our \Mdyn\ estimate is admittedly crude and its error  may be as large as a factor of 2 
even if the adopted inclination is correct.
If the rotation velocity increases toward the center at these scales, as we will see possible below, 
the dynamical mass is more likely to be overestimated.
For comparison, we estimated in our companion paper  
the mass of the central black hole in this galaxy to be $9\times 10^{6}$ \Msun\ 
\citep[for the distance of 34 Mpc]{Costagliola12}.
This value follows from the $K$ band bulge luminosity $10^{9.8} L_{\odot, K}$, which is calculated
from the total galaxy luminosity in Table \ref{t.4418param} and the mean bulge-to-total
luminosity ratio of 0.31 for Sa galaxies \citep{Graham08}.
The mass should have a factor of 2 uncertainty due to the scatter 
in the $L_{K, \rm bulge}$--$M_{\rm bh}$ correlation \citep{Marconi03}.

\begin{figure}[t]
\epsscale{0.45}
\plotone{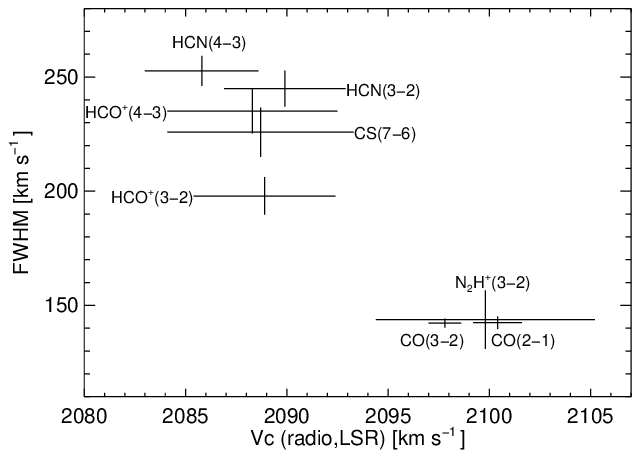} \\  
\plotone{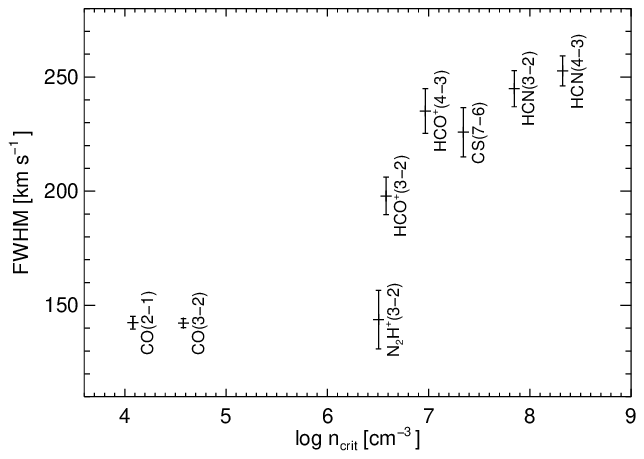}     
\caption{ \label{f.line-parameters}
(top) Line centroid velocity (\Vc) and FWHM for the lines in Table \ref{t.line_veloc}.
(bottom) Critical density (\ncrit) at 50 K and FWHM for the same lines.
Error bars are \plm4\signal\ for CO and  \plm1\signal\ for others.
Molecular parameters for the \ncrit\ are from the Leiden database \citep{LAMDA05}.
}
\end{figure}

We also infer from information of multiple lines that gas motion is faster toward the center.
Fig. \ref{f.line-parameters} and Table \ref{t.line_veloc} show
the centroid velocities and widths of the molecular lines measured in the SC configuration
data of \citest{Sakamoto10}.
An interesting observation here is that there are two groups 
in the centroid velocity -- line FWHM plane.
The first group consists of CO(3--2), CO(2--1), and \NtwoHplus(3--2) and have centroid velocities 
$\Vc \approx 2100$ \kms\ and FWHM $\approx$ 142 \kms.
The rest of the lines form the second group at $\Vc \approx 2088$ \kms\ and FWHM $\approx 234$ \kms.
Also, there is a trend that lines with higher critical densities are wider.
This variation of line parameters again suggests that the molecular gas in the center of NGC 4418 
has non-uniform properties or multiple components.
Indeed, the latter trend can be easily explained if the gas motion around the nucleus is 
faster closer to the center and if denser gas is more localized around the nucleus.
This model agrees with our previous observation in Table \ref{t.param} that
the lines with higher critical densities are more concentrated in the nucleus than CO.
If lines with higher critical densities indeed trace gas closer to the nucleus, 
and if we also assume that the gas distribution is axisymmetric around the nucleus,
the true systemic velocity of the nucleus would be about 2088 \kms.

\begin{figure}[t]
\begin{center}
\includegraphics[scale=1.0]{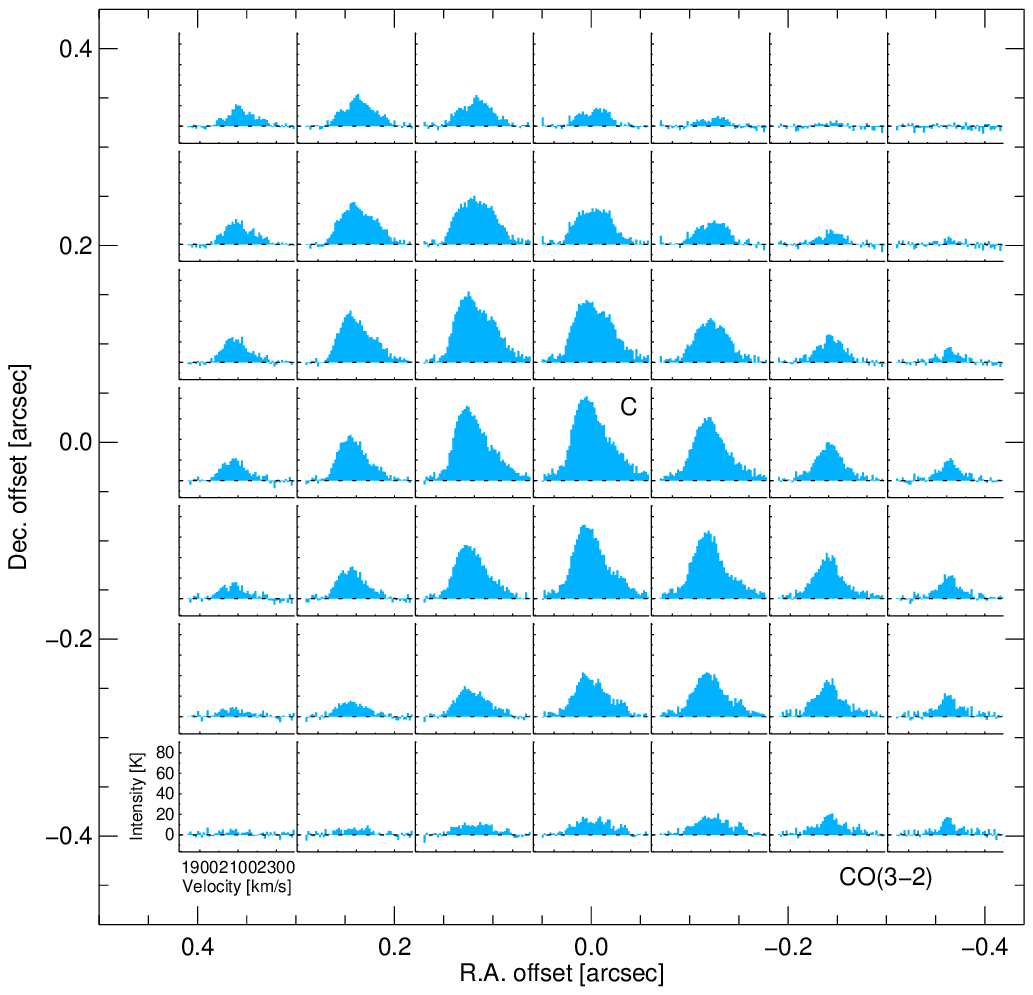} 
\includegraphics[scale=1.0]{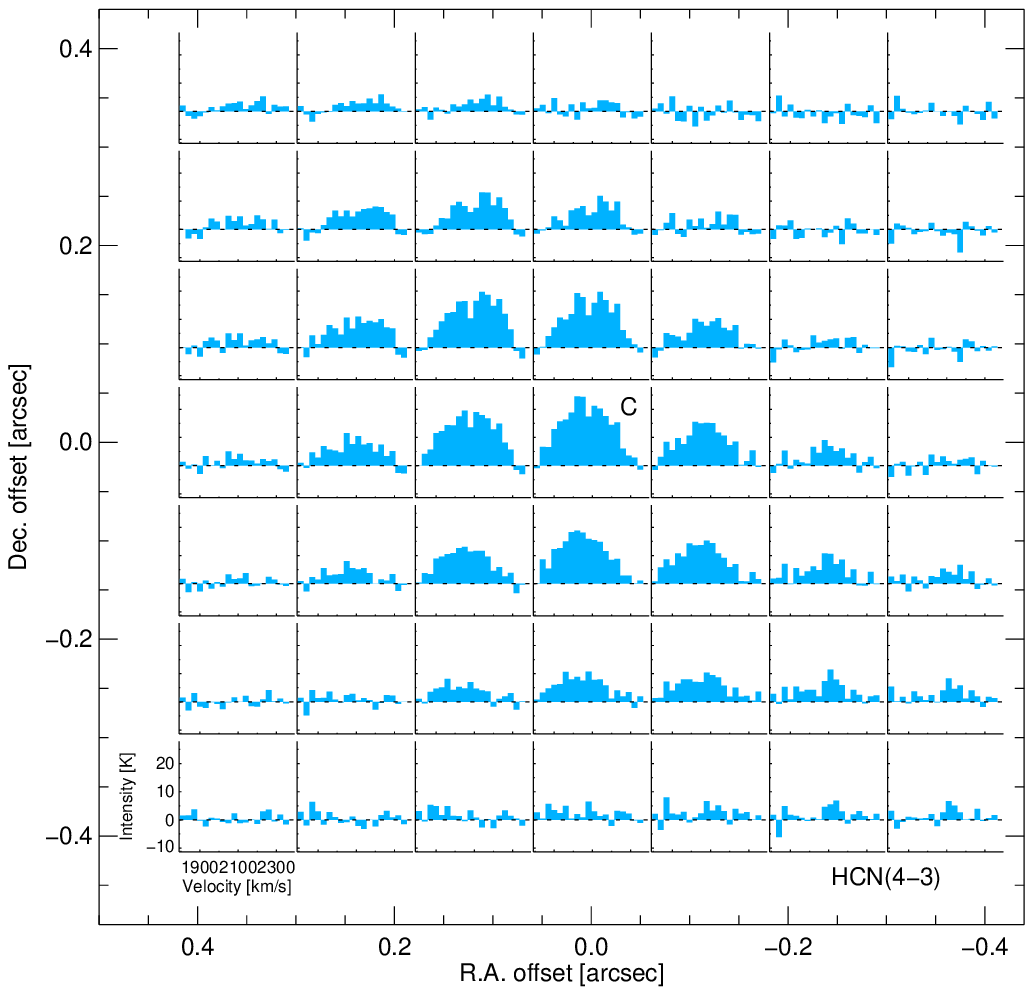} 
\end{center}
\caption{ \label{f.plcub}
CO(3--2) and HCN(4--3) line profiles in the central arcsecond of NGC 4418,
sampled on a grid of 0\farcs12 spacing.
The origin of the offset coordinates is at the 860 \micron\ continuum peak,
the spectrum observed at which is labeled with {\sf C}.
Velocity resolution of the spectra is 10 and 30 \kms\ for CO and HCN, respectively.  
The rms noise per channel is 2.4 and 2.2 K for CO and HCN, respectively.
}
\end{figure}

We can also evaluate radial motion of molecular gas from our data.
HCN spectra near the continuum peak in Fig.~\ref{f.plcub} and the HCN PV diagram
(Fig.~\ref{f.pv}) show a minor depression (2$\sigma$) at 2080 \kms.
This may be due to gas at about \Vsys\ and through line self-absorption 
or line absorption of the continuum from the core. 
The velocity of this possible absorption and the absence of clear absorption off the systemic velocity
indicate little radial motion of the HCN gas.
Meanwhile, the CO emission is significantly brighter below about 2100 \kms\ than at higher
velocities in the PV diagram. See also the CO line profile at the continuum core position
in Fig. \ref{f.plcub} where the line profile is asymmetric. 
If this is due to self-absorption of the CO emission, the foreground gas causing the absorption (and weaker
CO emission) is redshifted with respect to the systemic velocity and hence is flowing inward.
Although this asymmetry can be due to a chance imbalance of gas distribution between 
approaching and receding halves around the galactic center, this is qualitatively consistent
with the redshifted  \ion{O}{1}, OH, and \ion{H}{1} absorption observed
by \citet{Gonzalez-Alfonso12} and \citet{Costagliola12}.
The plausible gas motion mentioned here is opposite to that in Arp 220 where  
P-Cygni profiles of molecular lines suggested molecular outflows \citep{Sakamoto09}.

Gas motion outside the central 0\farcs5 does not appear to be ordinary rotation in our data.  
The CO mean velocity maps Figs.~\ref{f.SEVCOmom} (b) and \ref{f.VEXmaps} (b)
do not show an ordinary spider pattern of a rotating gas disk at $r \geq 0\farcs25$.
The velocity maps probably contain a mixture of rotational and radial motions and appear complex
because of non-uniform gas distribution and optical thickness of the CO line.

%%%%%%%%%%%%%%%%%%%%%%%%%
\subsection{Gas Properties}
\label{ss.gas_dust_properties}
Our high resolution observations also provide the following
new insights into gas  properties in the nucleus.

First, the CO(3--2) line has a peak brightness temperature of 90 K at 0\farcs3 resolution  (Table \ref{t.param}),
indicating warm molecular gas around the nucleus.
The brightness temperature of thermalized CO is the gas kinetic temperature diluted
by the CO opacity and
by the observing beam if it is larger than the source.
Both dilution effects are probably small for the extended \twelveCO\ in this case.
To put the observed \Tb\ into context, the peak value of 90 K at 50 pc resolution surpasses the peak CO brightness 
temperatures of about 50 K observed
in Arp 220 and NGC 253 at 100 pc (0\farcs3) and 20 pc (1\arcsec) resolutions, respectively \citep{Sakamoto08,Sakamoto11}.
Meanwhile both HCN(4--3) and \HCOplus(4--3) have peak brightness temperatures of about 30 K
at the same 0\farcs3 resolution  (Table \ref{t.param}). 
Their \about3 times lower \Tb\ than that of CO must be largely due to beam dilution effects
for these lines because they were found more compact than CO(3--2) in \S \ref{s.gas_dust_concentration}.
Subthermal excitation and lower opacities of HCN and \HCOplus\ may be additional reasons for
 the lower \Tb\ of these lines.

Second, the detection of lines from vibrationally excited molecules is another indication
of high temperatures in the nucleus. 
The HCN and \HCthreeN\ lines that we detected from their vibrationally excited states
have upper energy levels 510--1070 K above the ground level. 
Vibrational temperatures of these molecules have been estimated to be around 300 K
(\citealt{Costagliola10}; \citest{Sakamoto10,Costagliola12}).
Infrared radiation has been suggested to play a significant role for the vibrational excitation
in the \about100--200 K gas and dust that we inferred from brightness temperatures.
Our companion paper \citesp{Costagliola12} presents a more detailed analysis and discussion
on the excitation.

Third, the mean gas density  $\nHH \sim 5\times10^{5}$ \percubiccm\ of the \about20 pc core
(\S \ref{s.NH_Mgass}) is high enough for CO excitation
to J=3 but not for HCN and \HCOplus\ to J=4. 
Their critical densities for collisional excitation at 100 K are
\about$10^{4.5}, 10^{7}$, and $10^{8}$ \percubiccm, respectively.
The mean density is also far short of the \about$10^{11}$ \percubiccm\ critical density 
to vibrationally excite HCN with \HH\ collisions.
Therefore the excitation of the lines with high critical densities are probably achieved through photon
trapping, non-uniform density distribution to allow higher local densities than the mean,
mid-infrared radiation, and/or electron collisions \citesp{Sakamoto10,Costagliola12}.

Fourth, the ratios between lines and between line and continuum vary
with the area size of the measurements in the way expected when
the interstellar medium is warmer and denser at smaller radii.
This is a consequence of the size variation among various emission lines (\S \ref{s.gas_dust_concentration}).
For example the HCN(4--3)/CO(3--2) ratio of integrated brightness temperatures 
is $0.32\pm 0.03$ and $0.170 \pm 0.004$
at 0\farcs5 and 6\arcsec\ apertures, respectively. 
The ratio is expected to be higher when molecular gas is denser, warmer, or favorable for
HCN line emission in other ways.
Similarly, the equivalent widths of lines with respect to the continuum depend on the size scale.
The CO(3--2) equivalent width changes by a factor of 3, 
between $1.4\times10^{3}$ and $4.2\times10^{3}$ \kms, from 0\farcs5 to 6\arcsec\ scales.
The smaller equivalent width at the smaller scale can be explained by the higher opacity of the dust continuum 
at the continuum core \citesp{Sakamoto08}.
Finally, the ratio between HCN(4--3) to \HCOplus(4--3) is $2.08 \pm 0.27$ in the central 0\farcs5 while it
is $1.65 \pm 0.07$ in the central 6\arcsec\ \citesp{Sakamoto10}.
Among other differences, HCN has an order of magnitude higher critical density for collisional excitation 
and is also easier to radiatively excite than \HCOplus\ \citesp{Sakamoto10}. 
See \S \ref{s.implications} for more discussion on this line ratio.

%%%%%%%%%%%%%%%%%%%%%%%%%
\subsection{Optical Color Index}
\label{s.optical}
\begin{figure}[t]
\epsscale{0.7}
\plotone{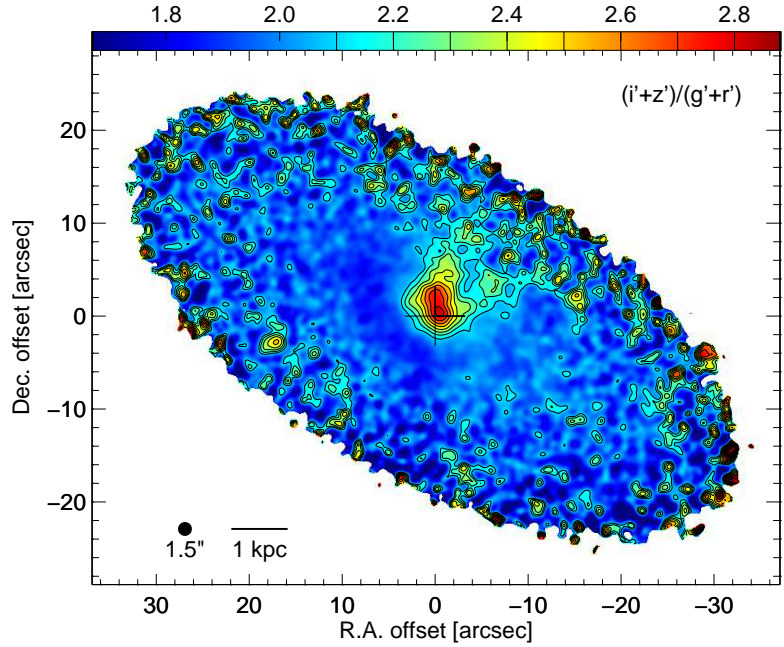} 
\caption{ \label{f.SDSS_color}
Distribution of optical color in the center of NGC 4418.
Flux ratio involving four SDSS bands, $(i' + z')/(g' + r')$, is plotted in such a way
that regions brighter in the longer wavelengths $i'$ and $z'$ than the shorter wavelengths $g'$ and $r'$
are shown in red while low ratio regions are shown in blue.
The resolution of this image is 1\farcs5.
The origin of the offset coordinates is at the 860 \micron\ continuum peak and is marked with a plus sign.
North is up and east is to the left.
}
\end{figure}

We found a U-shaped red feature in the multi-band optical images of the galaxy
from the Sloan Digital Sky Survey (SDSS).  
Fig. \ref{f.SDSS_color} shows the distribution of a flux ratio involving four SDSS bands, $(i'+z')/(g'+r')$.
Areas brighter at longer wavelengths ($i'$ and $z'$) are shown in red while
areas of lower flux ratios are in blue. 
The U-shaped red feature is along the northwestern semi-minor axis of the galaxy.
No counterpart is visible around the southeastern semi-minor axis.
The reddest point is almost on our 860 \micron\ continuum peak (i.e., the nucleus) 
and is about 0.5 mag (i.e., a factor \about1.6) redder than the off-center region outside the  
U-shape.
This feature extends at least up to 10\arcsec\ (1.7 kpc on the sky) from the nucleus.
The redder part of it near the nucleus has an elongation along the position angle of about 15\arcdeg,
similar to the faint CO(3--2) emission in Fig. \ref{f.SEVCOmom}.

We interpret the red feature to be an outflow cone emanating from the galactic center
along the rotation axis of the galaxy. 
The feature is very unlikely to be a structure on the galactic plane because such a
feature would be torn by differential rotation and also because the feature is on
the minor axis of the galaxy. 
The minor axis is determined by our viewpoint
while a radial feature in the galactic plane should have no preferred position angle with respect to us.
The outflow cone model does not have this problem 
because any feature along the rotation axis of the galaxy is always projected onto the minor axis. 
Also, the outflow can be bipolar in this model even though we only see half of it in the color index map.
The fact that we see the red feature only on the northwestern side suggests the side to be the near side of the outflow
and the far side of the galaxy disk, 
against which the outflow cone is visible as a red silhouette due to color-dependent extinction.
Dust in the cone probably causes the extinction and the reddening.
Our CO velocity data in Fig.~\ref{f.SEVCOmom} cover only a small fraction of the red optical feature
and do not reveal the gas motion in the red cone. 
The data only tell that the velocity field within a radius of \about500 pc
is more complicated than expected from purely circular rotation around the center.
However, \citet{Lehnert95} reported a 200 \kms\ velocity shift along the minor axis 
from their [\ion{N}{2}] line measurements for a survey of starburst winds.
The velocity shift suggests radial motion of the ionized gas along the polar axis.
Although we cannot yet confirm an {\it outward} radial motion
because the direction of the [\ion{N}{2}] velocity shift was not given, 
the velocity shift along the red cone better fits the outflow interpretation.
For this radial motion to be an inflow, gas and dust need to be falling in the U-shape cone toward the nucleus
along the polar axis, which seems too much of a coincidence.
Our outflow model predicts that the gas on the northwestern cone should be blueshifted on average with respect
to the nucleus.

Much closer to the nucleus, \citet{Evans03} found radial dark lanes out to about 3\arcsec\ from 
the nucleus in their near-infrared data. 
The lanes have an average position angle of about 15\arcdeg\ and are on both sides of 
the nucleus. 
The fainter lane to the north is in the same direction as the reddest part of the optical
U-shaped feature as well as the CO elongation.
There might be therefore a change of the outflow direction, first at P.A. $\sim$ 15\arcdeg\ in the
innermost region ($r< 3\arcsec \approx 500$ pc) and then along the minor axis of the galaxy (P.A. $\sim -30$\arcdeg)
outward.

We estimate the gas mass in the outflow to be  \about$4\times 10^{7}$ \Msol.
This is from the reddening and assumes a bipolar structure of the outflow.
The excess color index in $g' - r'$ is 0.8 mag at the central peak, 
about 0.5 mag in the reddest 1 kpc$^2$ near the base of the U-shape, 
and 1 mag kpc$^2$ when integrated over $2\times 2$ kpc$^2$ encompassing the entire U-shape.
Adopting the Milky-Way extinction law of $A_{g'} - A_{z'} = 0.71 A_{V}$ and
$\Sigma({\rm H+H_2})/A_{V} = 2 \times 10^{21}\; {\rm H\; \persquarecm} = 16\; \Msol\ \persquarepc$,
we estimate the gas column density to be  $N_{\rm H}=2 \times 10^{21}$ \persquarecm\ at the peak, 
the mean mass surface density in the reddest 1 kpc$^2$ to be about 10 \Msol\ \persquarepc,
and the gas mass in the one side of the outflow (i.e., the U-shape) to be about $2\times 10^{7}$ \Msol. 
The mass in the entire outflow is twice the last value on the assumption of a symmetric bipolar structure.
This mass estimate assumes foreground-screen extinction.
The 860 \micron\ dust core mixed with stars in the galactic center does not contribute much 
to the reddening and, appropriately, to the mass estimated here.
Note that the reddening data (and hence the outflow mass estimate from it) is more sensitive than
our CO data in which the U-shaped cone is not detected.
The cone would have a CO line brightness temperature on the order of only 0.1 K when the outflow
gas is 100\% molecular, its CO-to-\HH\ conversion factor is $2\times10^{20}$ \unitofX, 
and the line has a velocity width of 30 \kms\ at each position.
This is about 0.5$\sigma$ in our 0\farcs6 resolution CO(3--2) data cube.

We further estimate the kinematical age of the outflow to be on the order 10 $v_{200}^{-1}$ Myr 
and the gas outflow rate \about4 $v_{200}$ \Msol\ \peryr.
The parameter $v_{200}$ is the outflow velocity normalized by 200 \kms\ and is about unity
for the projected minor-axis velocity shear of 200 \kms\ measured in [\ion{N}{2}].
The age (i.e., crossing time) of the outflow is the ratio between the de-projected size of the outflow cone, 2 kpc, 
and the outflow velocity.
The mass outflow rate from the nucleus is from the gas mass above and the crossing time.
We caution about the uncertainties in these parameters due to
the unknown velocities of various outflow media, the poorly constrained extent of the flow,
and the uncertain projection effect. 
For example, the ionized gas traced by [\ion{N}{2}] and the dust causing the reddening may have different
bulk velocities. 
Also the opening angle of the outflow and the possible bend of the outflow direction add uncertainty
to our outflow velocity and to the parameters dependent on it.
For example, if the outflow velocity is as large as 1000 \kms\ seen in some molecular outflows \citep[e.g.,][]{Chung11,Sakamoto12}, 
the crossing time will be reduced to 2 Myr and the outflow rate will be \about20 \Msol\ \peryr.
Another caution is that, with the velocity information currently at hand, we cannot tell whether the gas and dust blown out 
from the nucleus to $\gtrsim 1$ kpc above the galactic plane will leave the galaxy or fall back to it.
In the latter case, the mass loss rate of the galaxy is smaller than the outflow rate from the nucleus.

%%%%%%%%%%%%%%%%%%%%%%%%%%%%%%%%%%%%%%%%%%%%%%%%%%%%%%%%%%%%
\section{Discussion}
\label{s.discussion}
%%%%%%%%%%%%%%%%%%%%%%%%%
\subsection{The Luminosity Source at the Nucleus}
We have obtained several parameters to constrain the nature of 
the luminosity source
such as the
\Lbol, \Mdyn, size, and  \NH\ of the luminous core,
and 
$\Sigma(\Lbol) \sim 10^{8.5}  \; \Lsol\ \persquarepc$
and $\Lbol/\Mdyn \sim 500$ \unitofLM\ calculated from them.
While we derived these from our 860 \micron\ observations alone they qualitatively agree with
the observations of \citet{Roche86} and many others that
most of the luminosity of the galaxy must come from a compact and deeply enshrouded nucleus.
The parameter values are comparable to or higher than those in  the ultraluminous infrared galaxy
Arp 220, where \citest{Sakamoto08} obtained 
$\Lbol/\Mdyn \gtrsim 400\; \unitofLM$ 
and $\Sigma(\Lbol) \gtrsim 10^{7.6}  \Lsol\; \persquarepc$ in the central 80 pc of the western nucleus.

%%%%%%%%%
\subsubsection{Constraint from $L/M$}
\label{s.constraint_LM}
\begin{figure}[t]
\epsscale{0.8}
\plottwo{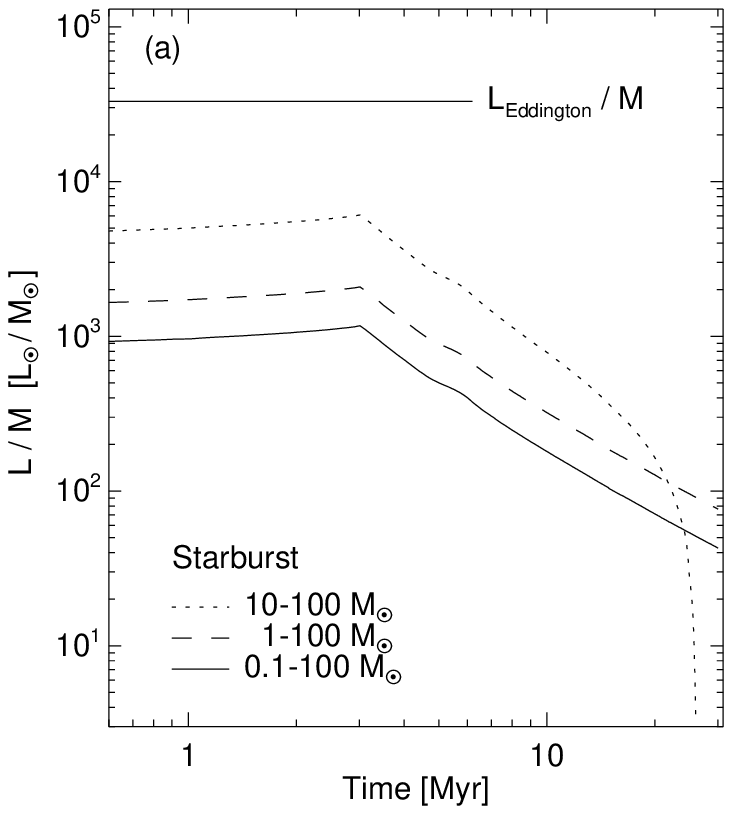}{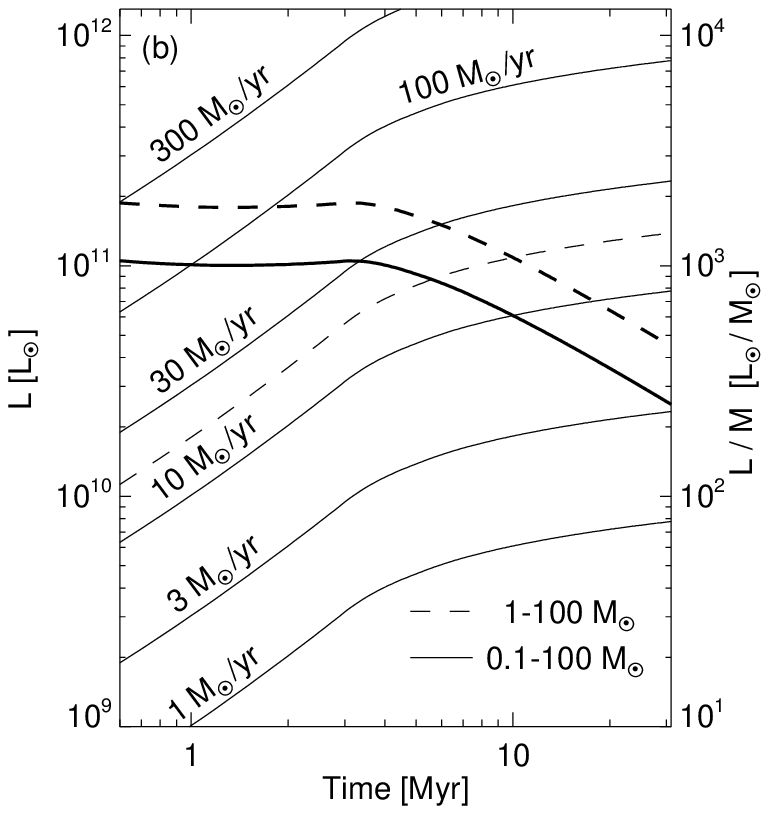} 
\caption{ \label{f.L_M}
Evolution of starburst parameters simulated using
{\sf Starburst99} 
and the IMF of \citet{Kroupa02}, which has
a power law index of $-1.3$ for 0.1--0.5 \Msol\ and $-2.3$ for 0.5--100 \Msol.
(a) 
Luminosity-to-mass ratios for single-age (i.e., instantaneous) starburst populations with three different IMF mass ranges.
Also plotted is the Eddington $L/M$ for fully ionized hydrogen plasma;
the mass is the black hole mass in the context of AGN.
(b)
Luminosity and luminosity-to-mass ratio of continuous starbursts with different SFRs.
Thin solid lines are starburst luminosities (left axis) for SFRs between 1 and 300 \Msun\ \peryear.
The thin dashed line is the starburst luminosity for SFR=10 \Msun\ \peryear\ and for the Kroupa IMF
truncated to the range of 1--100 \Msun.
The thick lines are the luminosity-to-mass ratios (right axis); the dashed one is for the same truncated IMF as above. 
The $L/M$ ratio is independent of the SFR.
}
\end{figure}

A young starburst can have a bolometric luminosity to mass ratio at or above
1000 \unitofLM\ for 3--10 Myr depending on the initial mass function
while a bare AGN at the Eddington luminosity has $L/\Mbh \approx 10^{4.5}$ \unitofLM.
Fig.~\ref{f.L_M} shows this for starbursts simulated using {\sf Starburst99} \citep{Leitherer99,Vazquez05,Leitherer10}
and the initial mass function (IMF) of \citet{Kroupa02}.
The duration of $L/M \gtrsim 1\times 10^{3} \; \Lsun\ \Msun^{-1}$ is about 5 Myr for the normal 
(i.e., untruncated) Kroupa IMF in both instantaneous and continuous starbursts.
In a continuous starburst, the combination of $L/M \approx 10^3$ \unitofLM\ and $L \approx 10^{11}$ \Lsol\ is obtained 
with the star formation rates (SFRs) of 30--100 \Msun\ \peryear\ at the ages of 3--1 Myr.
The starburst can be older and SFR lower if the IMF is biased toward high masses compared to the standard one
(e.g., age\about10 Myr and SFR\about10 \Msun\ \peryear\ for the IMF mass range of 1--100 \Msun).
The mass used for L/M in Fig. \ref{f.L_M} is that of newborn stars 
(or the black hole mass for the Eddington limit). 
The dynamical mass in our observed $\Lbol/\Mdyn$ includes
the gas covering the nucleus,
the central black hole, 
and the old stellar population preexisting the starburst or nuclear activity.
The gas mass is not negligible since it was estimated to be about 
half of the dynamical mass (\S \ref{s.NH_Mgass}).
Removing these from the denominator of the observed $L/M \sim 500$ \unitofLM\ will increase the ratio
to $\gtrsim$1000 \unitofLM. 
Fig.~\ref{f.L_M} therefore suggests that if a starburst is responsible for 
the core luminosity of NGC 4418 its $L/M$ is near the maximum 
that a starburst can have only when it is young, assuming the standard IMF.
The $L/M$ data from our submillimeter observations therefore provide a new way to
constrain the age in the starburst model.

Regarding whether a compact and massive star cluster mentioned above is able to form,
it has been argued that the maximum $L/M$ for a cluster-forming gas cloud or disk is 500--1000 \unitofLM\
\citep{Scoville03,Thompson05}.
This limit is because too much luminosity will blow out the star forming gas 
with radiation pressure on dust and halt the star formation.
The observed $L/M$ is close to this limit but does not clearly exceed it.
\citet{Murray09} predicted a mass-size relation for massive clusters taking the radiation pressure into account.
The diameters of $10^{7.5}$ and $10^{8}$ \Msol\ clusters, which can have $\sim$$10^{11}$ \Lsol\ when young,
are 5 and 10 pc respectively in his relation.
Thus at least these considerations do not reject the starburst model with the parameters we currently have.

If the energy source of the $10^{11} \Lsol$ core is an accreting black hole
then its Eddington ratio must be $\log (L/L_{\rm Edd}) = -0.5 \pm 0.3$
assuming the black hole mass of $10^{7.0 \pm 0.3}$\Msol\  (hence $L_{\rm Edd}=10^{11.5 \pm 0.3}$ \Lsol) 
estimated from the bulge luminosity (\citest{Costagliola12}, \S \ref{s.kinematics}).
Thus an AGN has no problem to be the dominant energy source as long as the $L/\Mbh$ is concerned.
It has no problem to be smaller than the 20 pc core either.

%%%%%%%%%%%%%%%%%%%%%%%%%%%
\subsubsection{Constraint from $\Sigma(\Lbol)$}
The luminosity surface density that we obtained, 
$\Sigma(\Lbol)=10^{8.5\pm 0.5} \; \Lsol\ \persquarepc$ in the central 20 pc (Table \ref{t.lbol}),
is among the highest compared to the  mid-infrared survey of 
(ultra)luminous infrared galaxies, starbursts, and Seyfert nuclei by
\citet{Soifer00,Soifer01,Soifer03,Soifer04}.
In their 0\farcs3--0\farcs5 resolution survey,
only three out of twenty-two nuclei were found to have
a surface brightness (or its lower limit) at $10^{8} \; \Lsol\ \persquarepc$ or larger.
All the three galaxies, Mrk 231, NGC 1275, and NGC 7469, have an AGN.
Their mid-IR cores have (upper limit) sizes of about 20 pc in two and 100 pc in one.
For comparison, the luminosity surface densities of star formation-dominated
nuclei are generally at or below $10^{7} \; \Lsol\ \persquarepc$.  
Between $10^{7}$ and  $10^{8}$ \Lsol\ \persquarepc\ are ultraluminous infrared galaxies where 
both AGN and starburst may be hidden behind a large amount of gas and dust  \citep{Soifer01}.
More recent observations by \citet{Imanishi11} also support
the mid-IR compactness and higher luminosity surface densities of AGNs compared to starbursts.
There is a theoretical model explaining the apparent limit for starbursts around $10^{7} \; \Lsol\ \persquarepc$.
In it a nuclear gas disk opaque at far-IR will have that luminosity surface density
when it supports itself by stellar radiation pressure and self-regulates
to have Toomre's $Q \sim 1$ \citep{Thompson05}. 
This is for dust temperatures less than about 200 K or for ``warm starbursts'' in \citet{Andrews11}.
For dust temperatures above 200 K, which the 0\farcs1 core of NGC 4418 may have, 
and for a gas surface density of $10^{5.8}$ \Msun\ \persquarepc\ that corresponds
to $\tau_{860\; \mu m} \sim 1$ of the continuum core, the disk at the Eddington limit  is in 
the ``hot starburst'' regime and will have a surface brightness of \about$10^{8.5} \; \Lsol\ \persquarepc$
\citep[][see their Equation 7]{Andrews11}.
According to our data and this model, the nucleus of NGC 4418 has either 
an energetically-dominant AGN or the hot starburst.

%%%%%%%%%%%%%%%%%%%%%%%%%%%
\subsubsection{Constraint from Mass Budget}
\label{s.MassBudget}
Our estimates of the masses in the central 30 pc of NGC 4418 are summarized in 
Table \ref{t.MassBudget}.
The mass of young stars there, $10^{8.0\pm0.5}$ \Msun, is for a model
young starburst having a luminosity of  $10^{11}$ \Lsol\  and a plausible luminosity-to-mass ratio of
$10^{3.0\pm0.5}$ \unitofLM. 
The mean star formation rate of this starburst would be 10--100 \Msol\ \peryr\
for a starburst age of 3 Myr. 
The mass of old stars in the central 30 pc, \about$10^{7.4}$ \Msun, 
is estimated for an average (\about$L^{*}$) Sa-type galaxy using
the K-band photometric parameters of \citet[their Table 3]{Graham08}, 
their conversion formulae to radial mass distribution \citep{Terzic05},
and 
the K-band mass to light ratio 1.06 \Msol/$L_{\odot\, K}$ for a 12 Gyr old stellar population with [Fe/H]=0
\citep{Worthey94}.
From the viewpoint of mass budget, AGN dominance in luminosity would eliminate the need for young stars
and help fit the sum of the component masses to the dynamical mass. 
The information currently at hand, however, does not let us conclude the AGN dominance from the mass budget 
because the masses used here typically have an uncertainty of a factor of a few.
The dynamical mass will be better constrained with higher resolution observations.

%%%%%%%%%%%%%%%%%%%%%%%%%%%
\subsubsection{Constraint from Time Scale}
The current tentative estimate of the outflow kinematical age, 10 Myr, is marginally 
longer than the age that an energetically-dominant starburst can have (\S \ref{s.constraint_LM}).
If this is firmly confirmed, it would be against the nascent starburst model.

%%%%%%%%%%%%%%%%%%%%%%%%%%%
\subsubsection{Constraint from \NH}
The gas shroud around the nucleus can shield the X-rays from the putative AGN
because it is highly Compton thick 
with $\NH \gtrsim 10^{25}$ \persquarecm\  to the galactic center (\S \ref{s.NH_Mgass}).

%%%%%%%%%%%%%%%%%%%%%%%%%%%
\subsubsection{Provisional Verdict}
\label{s.verdict}
To summarize on the luminosity source, the large luminosity of the nucleus can certainly be due to a
hidden AGN
as far as the parameters $L$, $L/M$,  $\Sigma(\Lbol)$, \NH, and mass budget of the nucleus are concerned.
It would have an Eddington ratio  \about0.3 with a factor of 2 uncertainty if the black hole and the bulge follow the
normal scaling relation.
This bolometric Eddington ratio is high compared to the median ratio of $10^{-3}$ among
local type 1 Seyferts \citep{Ho08}. 
In this sense the AGN mass accretion is rapid in NGC 4418 if the AGN is the dominant luminosity source.
The black hole should be growing at a higher rate than in most low luminosity AGNs.
The mass accretion rate to the black hole would be 0.1 \Msol\ \peryr\ for a 10\% radiative efficiency 
and the luminosity of $10^{11} \Lsol$ (i.e., $\Lbol \approx 0.1 \dot{M} c^2$). 
The characteristic time scale for the black hole growth is $10^{8}$ yr at this rate.
This AGN gas consumption rate is much lower than the star formation rate required for the
same luminosity because an AGN is more fuel-efficient.

A compact young starburst that has a mass \about$10^{8}$ \Msol, size $< 20$ pc, age \about\ a few Myr, and
$L/M \sim 10^3$ \unitofLM\ is an alternative that has not been excluded with our observations.
(The age could be up to 10 Myr as far as $L/M$ is concerned, but starbursts older than 3 Myr
are less favored because supernova explosions would be destructive to the dusty core as argued by \citet{Roussel03}.)
While such a starburst would be at an edge of the parameter space allowed for starbursts, 
it is noteworthy that the massive concentration of dense molecular gas is a favorable environment for star formation.
This is true even if there is a major AGN.

A combination of less luminous young starburst and a less luminous AGN is also possible
as long as their total luminosity is about $10^{11}$ \Lsol.
It remains to be seen with more accurate mass, luminosity, and size measurements and the constraints mentioned above
whether the nucleus of NGC 4418 is in the parameter region where only a dominant AGN is allowed.

%%%%%%%%%%%%%%%%%%%%%%%%%%%
\subsubsection{Consistency Check with Radio Data}
It is interesting to ask whether observations at radio wavelengths, which do not suffer from extinction by dust,
are consistent with an energetically dominant starburst in the nucleus.
As noted in \S \ref{s.constraint_LM}, a starburst consistent with our SMA observations has 
a star formation rate of about 30 \Msun\ \peryear\ in the last 3.3 Myr so that the starburst population
has a stellar mass of $10^{8}$ \Msun, a luminosity to mass ratio of $10^{3}$ \unitofLM,
and the observed luminosity of $\Lbol \sim 10^{11}$ \Lsun.
A younger starburst needs higher SFR to achieve the luminosity.
We note that the $L_{\rm IR}$--SFR relation of \citet{Kennicutt98} and \citet{Kennicutt12} gives 
SFR $\approx 15$ \Msun\ \peryear\ for the $10^{11} \Lsun$. 
However, as one can see in Fig.~\ref{f.L_M} (b),
a continuous starburst at that SFR will be about 10 Myr old when achieving 
the luminosity
and by that time the luminosity-to-mass ratio will drop by about 40\%
from its peak to \about600 \unitofLM, assuming the standard Kroupa IMF.
Thus, unless the IMF is truncated at low masses, our L/M constraint prefers higher SFRs than
estimated from the standard  $L_{\rm IR}$--SFR relation.

We estimate a lower limit of SFR $\gtrsim 9$ \Msun\ \peryear\ using
three constraints from the radio data and an SFR scaling relation of \citet{Murphy11}.
The first constraint is the high brightness temperature of the nucleus, 
$\gtrsim$$10^{5}$ K at \about2 GHz as reviewed in \S \ref{s.n4418review},
at the compact nuclear source (\about0\farcs1--0\farcs4).
This exceeds the temperature of \ion{H}{2} regions (\about$10^{4}$ K) and hence suggests 
the dominance of non-thermal synchrotron emission.
The second constraint is from the radio flux densities.
The nucleus has 1.49 GHz (20 cm) and 4.86 GHz (6 cm) flux densities of
38.5 mJy and 26.1 mJy, respectively, from a source with a $\lesssim 0\farcs4$ size \citep{Condon90,Baan06};
the resolutions of these observations are about 0\farcs4 and 0\farcs15, respectively.
The 20 cm luminosity of the nucleus is $5.3\times 10^{21}$ W \perHz\ assuming isotropy of the radio emission.
The third constraint is the spectral index between 20 and 6 cm, 
$\alpha_{\mbox{\small 20--6 cm}}=-0.33$ (for $S\propto \nu^\alpha$).
It is closer to the spectral index of thermal free-free emission ($\alpha=-0.1$)
than to that of non-thermal emission ($\alpha \approx -0.8$) 
both from optically thin star-forming regions \citep[e.g.,][]{Turner94,Murphy11}.
The apparent inconsistency between the first and third constraints
can be reconciled if the radio emission is partly opaque, i.e., if either synchrotron 
self-absorption or free-free absorption by thermal plasma is significant.
We therefore re-derived the 20 cm flux density using the 6 cm data
and a non-thermal spectral index of $-0.8$, while noting that 
this simplified correction will be insufficient if the 6 cm emission is also opaque.
The corrected 20 cm flux density and luminosity are 67.2 mJy and $9.3\times 10^{21}$ W \perHz, respectively.
The star formation rate of $\gtrsim 9$ \Msun\ \peryear\ is from
this 20 cm luminosity and the formula (14) of \citet{Murphy11}, which in our case is
$(\mbox{SFR/\Msun\peryr})=6.64\times10^{-22} (\nu/\mbox{GHz})^{-0.8} (L_\nu/\mbox{W\perHz})$.
This formula was derived using the proportionality of supernova rate to the star formation rate
(adopting the ratio between them from {\sf Starburst99})
and the empirical relation between the supernova rate and the non-thermal radio luminosity.
A similar star formation rate was derived in \citest{Costagliola12} from radio data albeit without the opacity correction.

The radio-derived SFR is smaller than needed for the starburst model at the face value.
This may be for the following (at least) two reasons {\it if} the starburst model is correct.
First, the radio SFR is a lower limit and can be higher as mentioned above.
Second, the SFR may be underestimated because the starburst is very young and is still deficient of supernovae
compared to older starbursts for which the SFR formula above was derived.
This nascent starburst is what \citet{Roussel03} argued for from the relative weakness of radio emission
compared to the infrared emission in this galaxy and by analyzing radio data in a way
similar to ours except that they used lower-resolution data (beam $\geq$ 45\arcsec).
An obvious weakness of this second explanation is that the dominance of non-thermal emission
(our first constraint) does not accord with the very young starburst.
The radio emission should be dominated by thermal emission from \ion{H}{2} regions 
rather than by non-thermal emission from supernovae-induced cosmic rays as long as the nucleus is
optically thin at radio wavelengths.
One could still attribute the weak thermal emission to direct absorption of ionizing photons by
dust and to high-opacity of the free-free emission. 
Non-thermal synchrotron emission can be less affected by these 
owing to the higher penetrating power of cosmic rays than ionizing photons.
These qualitative explanations are not particularly attractive because of their complexity, 
but we do not yet have a proof to rule them out.

In summary, the lower-limit SFR derived from radio data under the starburst-dominant model is 
about a factor of three smaller than the lower-limit SFR needed for the luminosity of the nucleus.
Although various uncertainties outlined above prevent us from excluding the starburst model, 
an energetically dominant starburst would need an elaborate model to explain the radio observations.

%%%%%%%%%%%%%%%%%%%%%%%%%%%
\subsubsection{Implications for Previous Studies}
\label{s.implications}
We also note implications of our new observations to a previous study of this energy source problem.
Full reanalysis of all existing observations in light of our new data is beyond the scope of this paper.

\citet{Imanishi04} used the presumed lack of shocks in the nucleus of this non-merging galaxy
and the high HCN-to-\HCOplus\ intensity ratio that they obtained toward the nucleus to infer
that the warm \HH\ they observed in the nucleus is due to X-ray heating. 
They then estimated the AGN luminosity from the \HH\ line luminosity and 
reached a conclusion that the AGN is the dominant energy source in the nucleus.
Our discovery of the red cone that we attribute to an outflow from the nucleus, as well as the
discovery by \citet{Gonzalez-Alfonso12} of the blue-shifted absorption presumably due to an inflow,  
raise the possibility of shock contribution to the infrared \HH\ lines.
The HCN-to-\HCOplus\ ratio does not give us the fraction of \HH\ luminosity due to an AGN.
Regarding the HCN-to-\HCOplus\ ratio itself,
while its enhancement around AGN gained support from more recent observations  at kpc resolutions \citep{Krips08},
it has been also found to vary within the center of a single galaxy from one giant molecular cloud to another  
in the range covered by the low-resolution observations of starbursts and AGNs \citep{Meier12}.
The latter authors suggested shock enhancement of HCN citing  stronger enhancement of HCN than \HCOplus\ 
in the molecular outflow from a protostar \citep{Bachiller97}.
In addition, for the particular case of NGC 4418, our detection and analysis of rotational lines from
vibrationally excited HCN toward the nucleus \citesp{Sakamoto10} further complicate the
line ratio analysis by introducing the excitation path through the vibrationally excited state. 
At present, no estimate exists for the fractional contributions of AGN X-rays and other mechanisms to 
the observed high HCN/\HCOplus\ ratio.
Thus these recent observations make the assumed 100\% contribution of X-ray heating to the \HH\ line luminosity 
uncertain. 
Therefore the AGN luminosity estimated from the \HH\ line luminosity is more uncertain than it was
at the time \citet{Imanishi04} interpreted their data.
This is not to say that the new observations are against the AGN hypothesis
because the AGN dominance can probably still explain the observations
and there are new observations that prefer an AGN for molecular chemistry in the nucleus \citep{GA12b}.
As was the case in the radio analysis reviewed above, we have a problem that
contributions of multiple (possible) mechanisms to observed properties make a conclusive data interpretation difficult.
It seems that we still need to accumulate more constraints to obtain
the breakdown of the luminosity sources in the nucleus.

%%%%%%%%%%%%%%%%%%%%%%%%%
\subsection{The Gas/Dust Shroud of the Nucleus and Comparison with Other Observations}
%%% 
The shroud of the $10^{11}$ \Lsol\ source is a \about20 pc core having $\tau_{\rm 860\, \mu m} \sim 1$ 
and the surrounding $\gtrsim$100 pc gas concentration that is generally warmer and denser toward the center.
The Compton-thick core has a surface temperature \about150 K and a mean density \about 5$\times10^{5}$ \HH\ \percubiccm\
(\S \ref{s.NH_Mgass}).
The radial variation of gas properties is indicated by 
the emission size sequence of  ``CO $>$ HCN, \HCOplus $>$ dust continuum'' (\S \ref{s.gas_dust_concentration}),
the continuum properties suggestive of ``dense core $+$ diffuse envelope'' structure (\S \ref{s.SED} and \ref{s.450cont}),
the larger velocity widths of lines with higher critical densities (\S \ref{s.kinematics}),
and
the dependence of line ratios and equivalent widths on area size (\S \ref{ss.gas_dust_properties}).
The same emission size sequence was observed in Arp 220 \citep{Sakamoto09}.
The declining temperature gradient from the center to larger radii 
has been an explanation for the deep 9.7 \micron\ depression in the mid-IR spectrum (\S \ref{s.n4418review}).

%%% 
Our observations on the properties of the central gas concentration
are generally consistent with those from the 1 mm observations in our companion paper
\citesp{Costagliola12}.
Leaving the comparison between the 1 mm and sub-mm observations
to that paper, we only mention that the deconvolved CO(2--1) peak brightness temperature 
in \citest{Costagliola12} is 80 K to agree with the CO(3--2) peak \Tb\ of 90 K
and that \citest{Costagliola12} also reports radial variation of gas properties.

\citet{Gonzalez-Alfonso12} made a detailed analysis of the circumnuclear
gas properties using Herschel spectroscopy at 50--200 \micron. 
They detected dozens of high-excitation lines, mostly in absorption, from \water, OH, \ion{O}{1}, HCN, 
\ammonia, and their \eighteenO\ isotopologues at 100--300 \kms\ resolutions.
Using these lines, 
they made a multi-component model in which the ISM is warmer in more compact components.
Our observations are consistent with this trend.
There are also agreements on the dust temperature around 150 K at $r \sim 10$ pc
and on the higher gas densities in more compact components down to the $r \sim 10$ pc component
(according to ($\tau_{\rm dust, 200\;\mu m}/{\rm size}$) in their Table 1).
A difference between their model and our observations is that we detected
about 0.2 Jy of 860 \micron\ continuum from a \about0\farcs1 (20 pc) core whereas
their model predicts a quarter of that from a component of the same size and about the
same temperature (150 K). 
Our larger flux density and brightness temperature lead us to a larger dust opacity for the core
at 860 \micron.
In order for the 50--200 \micron\ and 860 \micron\ observations to be consistent, 
it may be that the ISM in and around the core is not uniform to allow more leakage of emission 
from the warmer inner region than expected in the case of uniform shells.
Another possibility is that the dust $\beta$ is smaller than 2.
\citet{Gonzalez-Alfonso12} also obtained an oxygen isotope abundance comparable 
to that in the solar neighborhood, [\sixteenO/\eighteenO] \about 500.
As they noted, the lack of \eighteenO\ enhancement through stellar processing 
is in accord with the luminosity source
being an AGN and/or young pre-supernova starburst.

%%%%%%%%%%%%%%%%%%%%%%%%%%%%%%%%%%%%%%%%%%%%%%%%%%%%%%%%
\subsection{Evolution of the Luminous Nucleus}
The formation and evolution of the the hidden luminous nucleus
must be closely linked to those of the massive gas concentration at the nucleus.
It is because the concentrated gas affects the luminosity source and its appearance
through fueling and shielding.

\subsubsection{What drove the large amount of gas to the nucleus?}
\label{s.gas_concentration_mechanism}

The optical morphology of NGC 4418 (Fig.~\ref{f.sdss}) does not
show significant disturbance or distortion to suggest a major merger.
There is a blue companion \object{VV655} at 3\arcmin\ southeast (30 kpc on the sky) 
and at about the same redshift (2200 \kms)\footnote{What is cataloged in the atlas of interacting galaxies 
by \citet{VV77} is this companion itself and not the pair of VV655 and NGC 4418.}.
Although the two are probably physically associated with each other,
the large separation is unlike that in Arp 220 whose merger nuclei have a projected separation of only 0.3 kpc.
For Arp 220 there is no doubt that the concentration of a large amount of
gas to each nucleus was a consequence of the dynamical perturbation of the major merger.
In contrast, NGC 4418 does not appear to have acquired its nuclear gas concentration,
and hence the nuclear luminosity, through a major merger
although some properties of the nucleus are similar to those in Arp 220.

A stellar bar and minor merger are milder mechanisms for gas concentration than  
major merger and galaxy interaction. 
For example, stellar bars are known to transport cold gas in galaxy disks, sometimes as much as $>$$10^{8}$ \Msol,
to the galactic centers  \citep{Sakamoto99}.
If the gas concentration in NGC 4418 were due to a stellar bar
the situation may be similar to the one in the barred spiral galaxy \object{NGC 3504}
that has a single-peaked concentration of molecular gas 
with a scale length of 220 pc and a mass of $10^{8}$ \Msol\  in the central 200 pc \citep{Kenney93}.
The morphological classification of NGC 3504, (R)SAB(s)ab,  is very close
to that of NGC 4418 (Table \ref{t.4418param}) including the same moderate bar-type distortion.
Currently our data of NGC 4418 show neither a kinematical sign of bar-driven gas dynamics 
nor any clear sign of recent minor merger,
although such signs may be found in the future
in the somewhat disorderly gas motion in the central kpc (Fig. \ref{f.SEVCOmom}). 

The nuclear gas concentration may also be because most of the gas came from bulge stars. 
The gas originated from the galactic bulge through stellar winds and supernovae explosions 
would have little net angular momentum. 
Therefore some of it will sink to the galactic center when cooled.
A stellar population with a normal initial mass function returns about 30\% of its mass
to the interstellar space in $10^{10}$ yr \citep{Jungwiert01}.
The current bulge mass of NGC 4418 is about $0.6\times10^{10}$ \Msol, where we used the
K band bulge luminosity in \S \ref{s.kinematics} and the same luminosity to mass ratio as in \S \ref{s.MassBudget}.
The cumulative mass loss from the bulge stars should be therefore about $3\times10^{9}$ \Msol.
Stars formed from this gas should have small specific angular momentum and hence further mass loss from
them should keep the low specific angular momentum.
The \about$10^{9}$ \Msol\ molecular gas currently concentrated to the nucleus may be a fraction of the
low angular momentum gas originated this way from the bulge stars.
(See \citet{Young11}, \citet{Davis11}, and references therein for the general molecular gas content
in local early type galaxies and discussion on their origins.)

We note two more galaxies that have been recently found to host a compact and massive
concentration of molecular gas at their centers.
\object{NGC 1377} is an isolated lenticular galaxy 
that shares infrared characteristics such as the deep 9.7 \micron\ absorption with NGC 4418.
\citet{Aalto12} found a single-peaked \about100 pc-scale concentration of molecular gas 
at the nucleus and estimated a total gas mass in the central 400 pc of \about$10^{8}$ \Msol.
\object{NGC 1266} is another lenticular galaxy that has a strong concentration of molecular gas (\about$10^{8.5}$ \Msol) 
in the central 100 pc
without a clear sign of external disturbance or bar-driven gas transport \citep{Alatalo11}.
There may be a common mechanism to cause the compact massive gas concentrations
in these galaxies and NGC 4418. 
It may be the bulge-origin of the gas as mentioned above since these galaxies have large bulge fractions.

%%%%%%%%%%%%%%%%%%%%%%%%%%%%%%%%%%%%%
\subsubsection{Ongoing Evolution}
%%%  
Our discovery of an outflow feature from the center of NGC 4418 makes the nucleus
to have both an outflow and a previously known inflow. 
The inflow has been seen on our sightline near the galactic plane (\S \ref{s.kinematics}) 
and the outflow is perpendicular to the galactic plane (at least at $>$500 pc scales). 
Such complex gas motion has been modeled around a galaxy nucleus
hosting both starburst and an active nucleus  \citep[e.g.,][]{Wada02,Wada12}.
\citet{Ohsuga05} also simulated an outflow$+$inflow system at a smaller scale around a black hole
and showed that the system can achieve the Eddington luminosity or higher.
While we tentatively estimate the mass outflow rate to be \about4 \Msol\ \peryr,
\citet{Gonzalez-Alfonso12} estimated an inflow rate of $\lesssim 12 $\Msol\ \peryr\ from the redshifted
absorption lines they found.
The latter must be an overestimate because it assumed a spherically symmetric inflow without knowing the outflow.
It is therefore possible that the net outflow rate is close to zero or even positive.

The outflow from the deeply embedded nucleus may be driven at least at its root
by the radiation pressure on dust from the luminous nucleus \citep{Scoville03,Murray05}.
The criterion for a radiation-driven outflow is given with the luminosity-to-mass ratio as
\[
  \frac{\Lbol}{\Mdyn} > \frac{4 \pi c G}{\kappa},
\]
where $\kappa$ is the mass opacity coefficient, $c$ speed of light, and $G$ the gravitational constant.
For fully ionized hydrogen plasma, $\kappa$ is the ratio of the Thomson cross section to the masses of 
proton and electron and is $\sigma_{\rm T}/(m_{\rm p}+m_{\rm e})=0.4$  \unitofkappa.
The minimum $L/M$ for the outflow is thus the Eddington limit, $3.3\times10^4$ \unitofLM. 
For 3 \micron\ mass opacity coefficient of 50  \unitofkappa\ \citep{Scoville03}, where the opacity is due to dust,
the minimum $L/M$ for the outflow is 300 \unitofLM.
For 50 \micron\ where the SED of NGC 4418 peaks, the mass opacity coefficient is
$\kappa=2.5$ \unitofkappa\ \citep{Hildebrand83} and the minimum $L/M$ for the outflow is  5000 \unitofLM.
From these, a radiation-driven outflow appears possible from the region where the peak of the SED 
is in near-IR; the peak shifts to longer wavelength further from the central source. 
Unlike the inner region where the radiation pressure may drive the outflow, 
the outer region of the outflow along the rotation axis of the galaxy may be driven by
thermal pressure since the polar direction is the direction of the steepest pressure gradient.

With the inflow/outflow rates of several \Msol\ \peryr\ the ongoing evolution of 
the circumnuclear gas concentration is rapid 
in the sense that most of the \about$10^{8}$ \Msol\ molecular gas in the central 20 pc will be replaced in 10 Myr. 
Although the presumed accreting disk $+$ outflow structure may stay longer with a continuous supply of gas from the
larger gas reservoir outside the 20 pc core, the degree of nuclear obscuration may significantly vary in the gas-flow
time scale. 
The currently high $L/M$ can also change by an order of magnitude in 10 My for the starburst case
and by a larger factor in much shorter time in the AGN case.  
These considerations suggest a rapid evolution and a short duration of the current state
of the NGC 4418 nucleus.

Finally, NGC 4418 and the two galaxies mentioned in \S\ref{s.gas_concentration_mechanism} 
for their nuclear gas concentrations share an outflow of molecular gas and/or dust from each nucleus. 
The nuclear outflow from NGC 1377 is seen in CO kinematics \citep{Aalto12} and, as in NGC 4418, 
as a red cone on a semi-minor axis in an optical color-index image \citep{Heisler94}.
\citet{Alatalo11} found the outflow of NGC 1266 from observations of CO, \ion{H}{1}, and ionized gas 
and suggested it to be AGN-driven.
There may be a common evolutionary relation between the compact nuclear gas concentrations and the outflows 
in these galaxies because a gas concentration is generally expected to feed nuclear activities that can
drive an outflow.

%%%%%%%%%%%%%%%%%%%%%%%%%%%%%%%%%%%%%%%%%%%%%%%%%%%%
\section{Summary and Conclusions}
\label{s.summary}
We made high-resolution 860 and 450 \micron\ observations of NGC 4418, 
a luminous infrared galaxy having a deeply embedded nucleus.
Continuum and lines including CO(3--2), HCN(4--3), \HCOplus(4--3), and CS(7--6) in the 350 GHz band
were imaged at resolutions from  0\farcs3 to 1\arcsec(=165 pc)  in the central 34\arcsec\  (5.6 kpc).
Continuum at 660 GHz (450 \micron) was imaged at 0\farcs2 resolution.
We also examined SDSS optical images for circumnuclear features 
related to the luminous nucleus.
Our key results are the following:

1. 
The interstellar medium traced by the submillimeter emission is found strongly concentrated toward the nucleus.
CO(3--2) emission is detected only in the central 1 kpc (6\arcsec) of the galaxy with almost all the single-dish flux. 
More than half of the CO flux in the central kpc is in the central 165 pc (1\asec).
Higher excitation lines other than CO are more concentrated toward the nucleus 
(FWHM \about\  50 pc = 0\farcs3),
and the 860 \micron\ continuum emission is even more compact.

2.
The peak brightness temperature of CO(3--2) in the nuclear gas concentration is as high as 
90 K at  0\farcs3 (50 pc) resolution, indicating the presence of
hot molecular gas in the nucleus.
Other indications of hot or highly excited molecular gas 
are lines from vibrationally excited HCN and \HCthreeN\ with
upper energy levels at 660--1070 K above the ground.

3. 
The 860 \micron\ continuum emission at the nucleus  
has a size of \about0\farcs1 (17 pc) in FWHM and deconvolved peak brightness temperature (\Tb)
of about 210 K for a Gaussian model. 
The diameter is 27 pc and \Tb \about120 K for a uniform disk (sphere) model. 
The 860 \micron\ opacity of this core is on the order of 1 and therefore
the core is Compton thick ($\NH \gtrsim 10^{25} \; \persquarecm$).
The bolometric luminosity of this dusty core is estimated from these parameters 
to be $\Lbol \sim 10^{11.0}\; \Lsol$.
Most of the galaxy luminosity ($10^{11.1}$ \Lsol) is from this core.
Its luminosity surface density is $\Sigma(\Lbol) \sim 10^{8.5\pm 0.5} \; \Lsol\ \persquarepc$.

4.
Continuum emission at 450 \micron\ also peaks at the nucleus
but its flux density in our 30 pc ($\approx$0\farcs2) resolution data is
less than half of the total flux density from single-dish observations.
A more extended region than the \about0\farcs1 core, presumably the region
detected in CO(3--2), must have significant 450 \micron\ emission. 
Extinction of the core by this colder envelope is a likely additional reason 
for the small core contribution at 450 \micron.

5.
The dynamical mass within the central 30 pc is estimated to be \about$2\times 10^{8}$ \Msol\
from the molecular gas velocities in the central 0\farcs5.
The nucleus thus has $\Lbol/\Mdyn \sim 500  \; \Lsol/\Msol$.
Also, higher excitation lines are found broader than lower excited ones in the nucleus.
This is likely due to faster gas motion and higher gas excitation parameters (e.g., density and temperature)
toward the center. 
The CO line at $r = $0\farcs25-- 3\arcsec\ (40--500 pc) shows 
complex velocity structure that likely contains radial motion in addition to rotation.

6.
The observed high $L/M$, $\Sigma(L)$, and \NH\ as well as the nuclear mass budget
are consistent with a hidden AGN to be the main luminosity source
while they also allow a young starburst.
This confirms the proposition of \citet{Roche86} and subsequent researchers.
A dominant AGN would have a high Eddington ratio \about0.3 for the black hole expected from the bulge luminosity.
For star formation to achieve the observed $L/M$ and $\Sigma(\Lbol)$,  there must
be a compact concentration of $\lesssim 5$ Myr old stars well within the central 20 pc 
with a total mass on the order of $10^{8}$ \Msun.
Submillimeter data of higher quality will further constrain the nature of the nucleus.

7.
We found in the optical data a U-shaped red feature along a semi-minor axis of the galaxy.
We suggest it to be an outflow-induced cone seen in dust extinction (i.e., reddening).
We estimate the mass of the outflowing gas to be \about$4\times10^{7}$ \Msol\
and the rate of mass outflow from the nucleus to be \about4 \Msol\ \peryr, although the latter is tentative.
The luminous nucleus is blowing out its shroud in the polar direction 
while it is still fed from another direction as suggested from the previously known redshifted absorption.
Because these radial flows can replace most gas in the \about20 pc core in only $10^7$ yr,
they should play a significant role in the evolution of the luminous nucleus.

%%%%%%%
\acknowledgements
We thank the SMA staff for making these observations possible and Mark Gurwell for helping flux calibration.
KS thanks Dr. Eduardo Gonz{\'a}lez-Alfonso for answering questions about his recent work on NGC 4418.
This research made use of 
the NASA/IPAC Extragalactic Database (NED),
NASA's Astrophysics Data System (ADS),
the splatalogue database for astronomical spectroscopy,
the Sloan Digital Sky Survey (SDSS),
and
the JPL HORIZONS system. 
KS was supported by the Taiwanese National Science Council grant 99-2112-M-001-011-MY3.
SM acknowledges the co-funding of this work under the Marie Curie Actions of the European Commission (FP7-COFUND).

{\it Facilities:} \facility{SMA, Sloan}

%%%%%%%%%%%%%%%%%%%%%%%%%%%%%%%%%%%%%%%%%%%%%%%

\clearpage
%%%%%%%%%%%%%%%%%%%%%%%%%%%%%%%%%%%%%%%%%%%%%%%%%%%%%%%%%%%%%%%%
%%%%%%%%%%%%%%%%%%%%%%%%%%%%%%%%%%%%%%%%%%%%%%%%%%%%%%%%%%%%%%%%
% Tables.

%%%%%%%%%%%%%%%%%
% Table: galaxy parameters
\begin{deluxetable}{lcc}
\tabletypesize{\scriptsize}
\tablewidth{0pt}
\tablecaption{NGC 4418 parameters  \label{t.4418param} }
\tablehead{
	\colhead{Parameter}  &
	\colhead{Value}  &
	\colhead{note}
}
\startdata  
R.A. (J2000)         & 12\hr26\mn54\fs612       & (1) \\ 
Dec. (J2000)       & \minus00\arcdeg52\arcmin39\farcs41 & (1) \\
$V_{\rm sys}$ [\kms] &  2100 &  (2) \\
$D$ [Mpc] & 34 & (3) \\
Scale. 1\arcsec\ in pc & 165  \\
$L_{\rm 8-1000\mu m}$ [\Lsol] & $10^{11.1}$ & (4) \\
$L_{\rm K}$ [$L_{\rm \odot K}$] & $10^{10.3}$ & (5) \\
$L_{\rm B}$ [$L_{\rm \odot B}$] & $10^{9.8}$ & (6) \\
Hubble type & (R')SAB(s)a & (6) \\
P.A.  [\arcdeg] & 60 & (7) \\
$i$ [\arcdeg] & 62 & (8)
\enddata
\tablecomments{
(1) Position of the 860 \micron\ continuum nucleus measured in this work. The absolute positional uncertainty is
 estimated to be 0\farcs02 from the observations of a nearby quasar. 
Our pointing position was R.A.=12\hr26\mn54\fs620 Dec.=\minus00\arcdeg52\arcmin39\farcs40 taken from
the Sloan Digital Sky Survey Release 6 through the NASA Extragalactic Database.
(2) Centroid velocity of CO(2--1) spectrum \citesp{Sakamoto10}.
Its formal uncertainty from the Gaussian fit is less than 1 \kms\ but see \S \ref{s.kinematics}.
(3) The galaxy distance we adopt. 
For \Vsys(radio, LSR)=2100 \kms, 
the galaxy redshift corrected for local flow in the way of \citet{Mould00}
corresponds to an angular-size and luminosity distance of 33.5 and 34.1 Mpc, respectively,
for $H_0$=73 \kms\ \perMpc, $\Omega_{\rm matter}=0.73$, $\Omega_{\rm vacuum}=0.27$.
(4) From the IRAS flux measurements in \citet{Sanders03} and the formula in Table 1 of \citet{Sanders96}.
(5) Two Micron All Sky Survey, as listed in the NASA Extragalactic Database.
(6) \citet{RC3}.
(7) Position angle of the major axis from the Two Micron All Sky Survey Extended Source Catalogue 
\citep[2MASS XSC;][]{Jarrett00}.
(8) Inclination of the galaxy disk.
The minor to major axis ratio of $b/a=0.50$ in 2MASS XSC is translated 
to inclination assuming the intrinsic (edge-on) axial ratio of 0.2.
}
\end{deluxetable}

%%%%%%%%%%%%%%%%%
% Table: Obs. log
\begin{deluxetable}{clccllcccccc}
\tabletypesize{\scriptsize}
\tablewidth{0pt}
\tablecaption{Log of SMA observations  \label{t.obslog} }
\tablehead{ 
         \colhead{No.} &
	\colhead{UT date}  &
	\colhead{$f_{\rm LO}$} &
	\colhead{$N_{\rm ant}$} &	
	\multicolumn{2}{c}{array configuration} &
	\colhead{$L_{\rm baseline}$} &	
	\colhead{$\tau_{225}$} &
	\colhead{$\langle T_{\rm sys}\rangle$} &
	\colhead{$T_{\rm obs}$} 
	\\
	\colhead{ }  &	
	\colhead{ }  &
	\colhead{[GHz]}  &
	\colhead{ } &
	\colhead{name} &
	\colhead{pads} &
	\colhead{[m]} &	
	\colhead{} &
	\colhead{[K]} &
	\colhead{[hr]} 
	\\
	\colhead{(1)}  &
	\colhead{(2)}  &
	\colhead{(3)} &
	\colhead{(4)} &
	\colhead{(5)} &
	\colhead{(6)} &
	\colhead{(7)} &
	\colhead{(8)} &
	\colhead{(9)} &
	\colhead{(10)}	
}
\startdata
1 & 2009 Mar. 3 & 348.742 & 7 & VEX & 1,12,16,17,19,20,21   & 40--509  &  0.07 & 237 & 4.6  \\ 
2 & 2010 Mar. 3 & 347.698 & 6 & VEX & 1,16,17,18,19,20  & 85--509    & 0.04 & 153 & 4.3   \\  
$2'$ & 2010 Mar. 3 & 663.041 & 5 & VEX & 1,17,18,19,20  & 110--458 & 0.04  & 1228 & 3.8 \\    
3 & 2010 Mar. 28 & 347.698 & 6 & SC & 1,2,3,4,5,6  & 6--25    & 0.05 & 204 & 2.1  \\ 
4 & 2012 Feb. 5 & 347.728 & 7 & EXT & 1,9,12,14,15,16,17  & 24--222    & 0.04 & 163 & 4.9  
\enddata
\tablecomments{
(3) Frequency of the local oscillator.  
Upper (lower) sideband is from 4.0 to 6.0 GHz above (below) this frequency
for the first two tracks and from 4.0 to 8.0 GHz for the third and fourth tracks.
(4) Number of available antennas.
(5) SMA antenna configuration. VEX=very extended, EXT=extended, SC=sub-compact.
(6) Antenna locations. See \citet {Ho04} for a map with the numeric keys.
(7) Range of projected-length of baselines for NGC 4418.
(8) Zenith opacity at 225 GHz measured at the Caltech Submillimeter Observatory
adjacent to the SMA.
(9) Median double sideband (DSB) system temperature toward NGC 4418.
(10) Total integration time on the galaxy.
}
\end{deluxetable}

%%%%%%%%%%%%%%%%%
% Table: Data properties, resolutions and sensitivities of the cubes we used
\begin{deluxetable}{llccccccc}
\tablewidth{0pt}
\tablecaption{Data Properties  \label{t.data} }
\tablehead{ 
         \colhead{ID} &
	\colhead{Data Source}  &
	\colhead{SB.} &
	\colhead{Emission} &	
	\colhead{Beam} &
	\colhead{$\Delta V$} &	
	\multicolumn{2}{c}{Noise rms.} 
	\\
	\colhead{ }  &
	\colhead{ }  &	
	\colhead{ }  &
	\colhead{ }  &	
	\colhead{\asec $\times$ \asec} &
	\colhead{\kms} &
	\colhead{mJy \perbeam} &
	\colhead{K(R-J)} 
	\\
	\colhead{(1)}  &	
	\colhead{(2)}  &	
         \colhead{(3)}  &
	\colhead{(4)}  &	
	\colhead{(5)}  &	
	\colhead{(6)}  &	
	\colhead{(7)}  &	
	\colhead{(8)}  						         	
}
\startdata 
01 & 1     & U  & \HCOplus(4--3) etc.  & 0.34$\times$0.24  & 30  & 27  & 3.2  \\   
02 & 2     & U  & HCN(4--3) etc.          & 0.36$\times$0.26  & 30  & 20  & 2.2  \\   
03 & 1+2 & L     & CO(3--2)                 & 0.35$\times$0.26  & 10  & 22  & 2.4  \\  
04 & 1+2 & D    & 860 \micron\ cont.   & 0.35$\times$0.26  & \nd\tnm{a} & 2.4 & 0.27  \\       
05 & 1+3+4     & U  & \HCOplus(4--3) & 0.67$\times$0.45  &  20 & 15  & 0.47  \\   
06 & 2+3+4     & U  & HCN(4--3)          & 0.53$\times$0.41  &  20 & 16  & 0.72  \\   
07 & 1+2+3+4 & L  & CO(3--2)             & 0.69$\times$0.55  &  10 & 14  & 0.38  \\   
08 & 3+4          & L  & CS(7--6)              & 0.91$\times$0.59  &  20 & 16  & 0.32  \\  
09 & 3+4          & U & 4 GHz spectrum  & 0.92$\times$0.73  &  30 & 12  & 0.17 \\
10 & 3+4          & L & 4 GHz spectrum   & 1.00$\times$0.76  &  30 & 11  & 0.15 \\
11 & $2'$         & D & 450  \micron\ cont.   & 0.23$\times$0.15  & \nd\tnm{a} & 65 & 5.3  
\enddata
\tablecomments{
Column (2): Data source. The numbers are those in Table \ref{t.obslog}. 
Column (3): Sideband. U=USB, L=LSB, D=double sideband=USB+LSB.
Column (4): Main emission in the data.
Column (5): Major and minor axes (FWHM) of the synthesized beam. 
Column (6): Velocity resolution.
Columns (7) and (8): Noise r.m.s. intensity in the data. For lines this is for a channel of the width in (6).
The noise in (8) is in kelvin and in Rayleigh-Jeans brightness temperature. 
}
\tablenotetext{a}{The mean frequency and total bandwidth of the continuum are
respectively 348.2 and 3.5 GHz for the 860 \micron\ data and 663.0 and 3.9 GHz for the 450 \micron\ data.  }
\end{deluxetable}

%%%%%%%%%%%%%%%%%
% Table: peak Tb and flux in the VEX data
\begin{deluxetable}{lcccc}
\tablewidth{0pt}
\tablecaption{Peak brightness temperatures and fluxes  in the VEX data \label{t.param} }
\tablehead{ 
	\colhead{Emission}  &
	\multicolumn{2}{c}{max \Tb} & 	
	\colhead{$S$, $S_\lambda$ (0\farcs5 FWHM)} &
	\colhead{$f(0\farcs5/6\asec)$} 		
	\\
	\colhead{ }  &
	\colhead{K(R-J)} &
	\colhead{K(Planck)} &	
	\colhead{Jy \kms, mJy } & 			
	\colhead{ }  		
	\\
	\colhead{(1)}  &	
         \colhead{(2)}  &
	\colhead{(3)}  &	
	\colhead{(4)}  &	
	\colhead{(5)} 		
}
\startdata  
CO(J=3--2)                                   & 82 & 90  & $240\pm 3$    & $0.32\pm0.02$       \\
HCN(J=4--3)                                & 24 & 31  & $80\pm 3$      & $0.60\pm0.05$       \\
\HCOplus(J=4--3)                        & 20 & 28 & $39\pm 4$      & $0.48\pm0.06$       \\
\HthirteenCN(J=4--3)                  & 13 & 20 & $29\pm 4$      & $0.71\pm0.13$       \\
\HCfifteenN(J=4--3)                    &  8 & 15   & $17\pm 4$      & $1.55\pm0.57$       \\
HCN($v_2$=$1^{1f}$,J=4--3)    & 13 & 20 & $18\pm 4$      & $0.60\pm0.15$        \\
860 \micron\ continuum             & 17 & 24 & $168 \pm 3$   &  $0.95\pm0.07$      \\
450  \micron\ continuum            & 31 &  45 & $369 \pm 37$   &  \nd      
\enddata
\tablecomments{
Columns (2) and (3): 
Peak brightness temperatures measured in the VEX data (data ID 01--04 and 11 in Table \ref{t.data}).
No correction is made for missing flux and beam dilution.
Temperatures calculated using the Rayleigh-Jeans approximation and the Planck function are given in (2) and (3), respectively.  
The $\max  \Tb$  is measured in the 10 \kms\ resolution data for CO and 30 \kms\ for other lines.
Column (4):  
Line fluxes and continuum flux densities that we detected in the central 0\farcs5 (FWHM).
They were measured from data ID 01--04 and 11 in Table \ref{t.data} after convolving them to the 
resolution.
The line fluxes are integrated over 1900--2300 \kms; for \HthirteenCN\ and \HCfifteenN\ we doubled the
flux of half that width since the other half of the line is either outside of our bandwidth or blended with another line.
Note that CO, HCN, and  HCN($v_2$=1) lines are blended with a nearby \HCthreeN\ transition.
The 1$\sigma$ errors are for the thermal noise. 
We estimate additional flux-scaling error to be \about5 \% for 345 GHz
data and 40 \% for 660 GHz data.
Column (5): 
Fraction of flux detected in our 0\farcs5 data compared to the 6\arcsec\ resolution data in \citest{Sakamoto10} .
}
\end{deluxetable}

%%%%%%%%%%%%%%%%%
% Table: Vis. fitting result
\begin{deluxetable}{lcccc}
\tablewidth{0pt}
\tablecaption{Visibility Fitting Results  \label{t.visfit} }
\tablehead{
	\colhead{source}  &
	\colhead{$\lambda$, line}  &
	\colhead{flux density}  &
	\colhead{size}  &	
	\colhead{reduced $\chi^2$}  
         \\
	\colhead{ }  &
	\colhead{mm}  &
	\colhead{Jy}  &
	\colhead{arcsec}  &	
	\colhead{ } 
         \\
	\colhead{(1)}  &
	\colhead{(2)}  &
	\colhead{(3)}  &
	\colhead{(4)}  &	
	\colhead{(5)}  
}
\startdata
NGC 4418      & 0.85       & 0.205 $\pm$ 0.002 & 0.101 $\pm$ 0.011  & 0.63  \\
NGC 4418      & HCN(4--3)  & 0.299 $\pm$ 0.006 & 0.245 $\pm$ 0.012  & 2.3  \\
NGC 4418      & \HCOplus(4--3)  & 0.204 $\pm$ 0.006 & 0.343 $\pm$ 0.028  & 0.51  \\
Titan                & 0.85      & 3.41 $\pm$ 0.08 & 0.848 $\pm$ 0.008  & 1.9  \\ 
Titan                & 0.88       & 3.28 $\pm$ 0.06 & 0.863 $\pm$ 0.007  & 1.9  \\
J1222+042     & 0.85      & 0.55 $\pm$ 0.02 & 0.049 $\pm$ 0.023  & 2.8   \\
J1222+042     & 0.88       & 0.53 $\pm$ 0.01 & 0.0 $\pm$ 0.0           & 3.2  
\enddata                     
\tablecomments{
Fitting of NGC 4418 used data from multiple array configurations;
2010 VEX, SC, and 2012 EXT. The VEX data of \HCOplus\ had too low S/N to be usable.
Fitting of Titan and the quasar J1222+042  were to verify our data analysis.
The size of Titan based on ephemeris is 0\farcs8425.
Column (2):
Continuum wavelength or line name. 
The 0.85 mm continuum is from USB and 0.88 mm is from LSB.
Column (3): 
Flux density at zero baseline (i.e., total flux density).
For lines, this is for the line emission averaged from 1900 to 2300 \kms.
Column (4):
The size is FWHM of a Gaussian for quasars and NGC 4418, and the diameter of
a uniform-brightness disk for Titan.
}
\end{deluxetable}

%%%%%%%%%%%%%%%%%
% Table: Lbol of the nucleus
\begin{deluxetable}{ccccccc}
\tablewidth{0pt}
\tablecaption{Luminosity Estimates of the Nucleus  \label{t.lbol} }
\tablehead{
	\colhead{shape}  &
	\multicolumn{2}{c}{size} &
	\colhead{\Tb} &
	\colhead{$\log \Lbol$} &
	\colhead{$\log \Sigma(\Lbol)$} &
	\colhead{$\log \rho(\Lbol)$} 	
	\\
	\colhead{ }  &
	\colhead{mas}  &
	\colhead{pc} &
	\colhead{K} &
	\colhead{\Lsol}	 &
	\colhead{\Lsol\ \persquarepc} &
	\colhead{\Lsol\ \percubicpc} 
	\\
	\colhead{(1) }  &
	\colhead{(2)}  &
	\colhead{(3)} &
	\colhead{(4)} &
	\colhead{(5)} &
	\colhead{(6)} &
	\colhead{(7)}		
}
\startdata
Gaussian       & $101\pm11$ & $17\pm2$ & $205 \pm 47$ & $11.3 \pm 0.3$ & $9.0 \pm 0.4$ & $8.0 \pm 0.4$ \\
Disk/Sphere  & $162\pm18$ & $27\pm 3$ & $119 \pm 27$ & $10.8\pm 0.3$ & $8.1 \pm 0.4$ &  $6.8 \pm 0.4$
\enddata
\tablecomments{
Column (1): 
Assumed shape of the source to calculate the peak brightness temperature. 
Disk is a circular disk (i.e., a sphere projected on the sky) with uniform surface brightness.
Columns (2) and (3): 
Source size is the FWHM for the Gaussian model and the diameter for the Disk/Sphere model.
Column (4): 
Brightness temperature calculated with the Planck function. 
This is the peak value for the Gaussian model and the value across the surface in the Disk/Sphere model.
Column (5): 
Bolometric luminosity of the source calculated from the source size, temperature, and the 
Stefan-Boltzmann law. 
See \S \ref{s.Tb_Lbol}.
Columns (6) and (7):
The mean surface density and volume density of the luminosity within the size in the column (3).
}
\end{deluxetable}

%%%%%%%%%%%%%%%%%%%%%%
% Table: line velocities and widths
\begin{deluxetable}{lcc}
\tablewidth{0pt}
\tablecaption{Line Velocities and Widths  \label{t.line_veloc} }
\tablehead{ 
	\colhead{Line}  &
	\colhead{$V_{\rm c}$} &
	\colhead{FWHM} 			
	\\
	\colhead{ }  &
	\colhead{\kms} &		
	\colhead{\kms} 		
}
\startdata  
\HCOplus(4--3)                        &    2088.3 \plm\ 4.2  & 235.1 \plm\ \phn9.8   \\
HCN(4--3)                                 &    2085.8 \plm\ 2.8  & 252.7 \plm\ \phn6.6   \\
CO(3--2)                                    &    2097.8 \plm\ 0.2  & 142.2 \plm\ \phn0.5   \\
CS(7--6)                                    &    2088.7 \plm\ 4.6  & 225.8  \plm\ 10.8        \\
\NtwoHplus(3--2)                     &    2099.8 \plm\ 5.4  & 143.8  \plm\ 12.8        \\
\HCOplus(3--2)                        &    2088.9 \plm\ 3.5  & 197.9 \plm\ \phn8.2   \\
HCN(3--2)                                 &    2089.9 \plm\ 3.0  & 244.9 \plm\ \phn7.9   \\
CO(2--1)                                    &    2100.4 \plm\ 0.3  & 142.4 \plm\ \phn0.7   
\enddata
\tablecomments{
Centroid velocities and widths of molecular lines in the central 6\arcsec\ of NGC 4418.
The velocities are in the radio definition and with respect to the LSR.
These parameters are from Gaussian fitting to the spectra in \citest{Sakamoto10}.
Errors are \plm1$\sigma$.
HCN(4--3) is blended with a \HCthreeN\ line on the low velocity side. 
}
\end{deluxetable}

%%%%%%%%%%%%%%%%%%%%%%
% Table: mass budget of the nucleus
\begin{deluxetable}{lcc}
\tablewidth{0pt}
\tablecaption{Mass Budget in the Central 30 pc\label{t.MassBudget} }
\tablehead{ 
	\colhead{type}  &
	\colhead{$\log M$} &
	\colhead{note} 			
	\\
	\colhead{ }  &
	\colhead{[\Msol]} &		
	\colhead{} 		
}
\startdata  
$\Mdyn(r\leq {\rm 15\; pc})$ & $8.3\pm0.3$   & (a)  \\
\Mbh                                         & $7.0 \pm 0.3$ & (a) \\
$\Mmol(r\leq {\rm 10\; pc})$ & \about$8.0$    &  (b)\\
$M_{\rm young\,stars}(r\leq {\rm 10\; pc})$ & $8.0\pm0.5$  & (c) \\
$M_{\rm old\,stars}(r\leq {\rm 15\; pc})$ & \about$7.4$         & (d)
 \enddata
\tablecomments{Numbers with \about\ may well have 0.3--0.5 dex uncertainties as others.}
\tablenotetext{a}{See \S \ref{s.kinematics}.}
\tablenotetext{b}{See \S \ref{s.NH_Mgass}.}
\tablenotetext{c}{See \S \ref{s.MassBudget}.
Most of the luminosity-bearing young stars must be well within the central 20 pc. }
\tablenotetext{d}{See \S \ref{s.MassBudget}.
For an average Sa galaxy with $M_K=-23.8$ mag. NGC 4418 is 0.3 times
as luminous.}
\end{deluxetable}

%%%%%%%%%%%%%%%%%%%%%%%%%%%%%%%%%%%%%%%%%%%%%%%%%%%%%%%%
%%%%%%%%%%%%%%%%%%%%%%%%%%%%%%%%%%%%%%%%%%%%%%%%%%%%%%%%
% Figures.
\clearpage

%%%%%%%%%%%%%%%%%%%%%%%%%
% SDSS image of N4418

%%%%%%%%%%%%%%%%%%%%%%%%%
% N4418 VEX345 spectra

%%%%%%%%%%%%%%%%%%%%%%%%%
% N4418 SubExt345 spectra

%%%%%%%%%%%%%%%%%%%%%%%%%
% N4418 Sub+EXT+VEX CO(3-2) channel maps.

%%%%%%%%%%%%%%%%%%%%%%%%%
% N4418 VEX CO(3-2) channel maps.

%%%%%%%%%%%%%%%%%%%%%%%%%
% N4418 VEX HCN(4-3) channel maps.

%%%%%%%%%%%%%%%%%%%%%%%%%
% N4418 CO(3-2) low-res. (0.6") moment maps from Sub+EXT+VEX data.

%%%%%%%%%%%%%%%%%%%%%%%%%
% N4418 VEX345,660 moment maps.

\clearpage
%%%%%%%%%%%%%%%%%%%%%%%%%

%%%%%%%%%%%%%%%%%%%%%%%%%
% SED

%%%%%%%%%%%%%%%%%%%%%%%%%
% Fraction of Flux Recovery in our 660 GHz data

%%%%%%%%%%%%%%%%%%%%%%%%%
% N4418 Sub+EXT+VEX HCN/HCO+/CS moment maps

%%%%%%%%%%%%%%%%%%%%%%%%%
% VEX PV maps

%%%%%%%%%%%%%%%%%%%%%%%%%
% line Vc - FWHM

%%%%%%%%%%%%%%%%%%%%%%%%%
% VEX profile maps. CO and HCN

%%%%%%%%%%%%%%%%%%%%%%%%%
% SDSS optical image

%%%%%%%%%%%%%%%%%%%%%%%%%
% L/M of stars with various low-mass cutoff and ages

\end{document}